\documentclass[
superscriptaddress,
 amsmath,amssymb,
 aps,
longbibliography
]{revtex4-2}

\usepackage{graphicx}
\usepackage{dcolumn}
\usepackage{bm}
\usepackage{hyperref}
\usepackage[normalem]{ulem}

\newcommand{\dtPTildePrime}{\ensuremath{\frac{\partial \tilde{p}'}{\partial \tilde{t}}}}
\newcommand{\dxPTildePrime}{\ensuremath{\frac{\partial \tilde{p}'}{\partial \tilde{x}}}}
\newcommand{\dtUTildePrime}{\ensuremath{\frac{\partial \tilde{u}'}{\partial \tilde{t}}}}
\newcommand{\dxUTildePrime}{\ensuremath{\frac{\partial \tilde{u}'}{\partial \tilde{x}}}}

\newcommand{\rev}{\textcolor{black}}

\usepackage{tikz}
\usetikzlibrary{positioning}
\usetikzlibrary{arrows.meta}
\usepackage{printlen}
\newlength\myboxwidth
\begin{document}

\title{Hard-constrained neural networks for modelling  nonlinear acoustics}
\author{Defne E. Ozan}
\affiliation{Department of Aeronautics, Imperial College London, SW7 2AZ, London, UK}
\author{Luca Magri}
\email{l.magri@imperial.ac.uk}
\affiliation{Department of Aeronautics, Imperial College London, SW7 2AZ, London, UK}
\affiliation{The Alan Turing Institute, NW1 2DB, London, UK}

\begin{abstract}
In this computational paper, we model acoustic dynamics in space and time from synthetic sensor data. The tasks are (i) to predict and extrapolate the spatiotemporal dynamics, and (ii) reconstruct the acoustic state from partial observations. To achieve this, we develop acoustic neural networks. These are networks that learn from sensor data, whilst being constrained by prior knowledge on acoustic and wave physics. The prior knowledge is constrained as \rev{a} soft constraint, which informs the training, and as a hard constraint (Galerkin neural networks), which constrains parts of the network's architecture as an inductive bias. First, we show that standard feedforward neural networks are unable to extrapolate in time, even in the simplest case of periodic oscillations. This motivates the constraints on the prior knowledge. Second, we constrain the prior knowledge on acoustics in increasingly effective ways by  (i) employing periodic activations (periodically activated neural networks); (ii) informing the training of the networks with a penalty term that favours solutions that fulfil the governing equations (soft-constrained); (iii)  constraining the architecture in a physically-motivated solution space (hard-constrained); and (iv) combination of these. Third, we apply the networks on two testcases for two tasks in nonlinear regimes, from periodic to chaotic oscillations. The first testcase is a twin experiment, in which the data is produced by a prototypical time-delayed model. In the second testcase, the data is generated by a higher-fidelity model with mean-flow effects and a kinematic model for the flame source. We find that (i) constraining the physics in the architecture improves interpolation whilst requiring smaller network sizes, (ii) extrapolation in time is achieved by periodic activations, and (iii) velocity can be reconstructed accurately from only pressure measurements with a combination of physics-based hard and soft constraints. In acoustics and thermoacoustics, this works opens possibilities for physics-constrained data-driven modelling. Beyond acoustics, this work opens strategies for constraining the physics in the architecture, rather than the training.

\end{abstract}

\maketitle

\section{Introduction}
When modelling, reconstructing, and forecasting dynamics from data, constraining prior knowledge into machine learning methods can significantly improve prediction, robustness and generalizability \citep[e.g.,][]{duraisamy2019turbulence,brunton2020machine,Karniadakis2021,doan2021short}. 
On the one hand, constraints that are imposed in the loss function as penalty terms, which act during training, are referred to as ``soft constraints". Soft constraints can include the governing partial differential equations (PDEs) derived from conservation laws, initial and boundary conditions \citep{lagaris1998artificial}. Deep feedforward neural networks with soft-constrained physics information, coined as physics-informed neural networks (PINNs)~\citep{Raissi2019}, have been employed to infer flow fields from synthetic data from prototypical flows \citep{Raissi2019_sin,raissi_hidden_2020}, from puffing pool fires \citep{sitte_velocity_2022}, experimental data from a flow over an espresso cup \citep{cai_flow_2021}, and clinical MRI data \citep{kissas_machine_2020}, to name only a few. Beyond PINNs, physics information has enabled super-resolution tasks without high-resolution labels in deep feedforward neural networks \citep{fathi_super-resolution_2020, wang_dense_2022, Eivazi2022} and in convolutional neural networks \citep{gao_super-resolution_2021, Kelshaw2022}. 
On the other hand, constraints that are imposed in the architecture (as opposed to the training) are referred to as ``hard constraints". Hard constraints span the areas of  the function space in which physical solutions live, i.e., they create an {\it inductive bias} \citep{goyal_inductive_2022}. Known PDEs \citep{chen_theory-guided_2021,xu_physics_2022,mohan_embedding_2023}, invariances \citep{ling_reynolds_2016}, Dirichlet boundary conditions \citep{lagaris1998artificial}, and periodic boundary conditions \citep{Zhang2020,Dong2021,Kelshaw2022} have been incorporated in the architecture of neural networks as hard constraints. 
In this paper, we design Galerkin neural networks, which hard-encode the solution structure into the architecture. These networks are inspired by Galerkin projection, which is a common technique for solving PDEs by projecting the equations onto a finite set of modes, which transforms the problem into a set of  ordinary differential equations \citep{HolmesGalerkin}. The library of modes can be a generic basis such as Fourier, or data-driven such as proper orthogonal decomposition (POD) modes. In this paper, we take advantage of a physical basis.
%
%

We focus on solutions from acoustics and thermoacoustics, which originate from nonlinear wave equations. Thermoacoustic systems contain nonlinearities in the \rev{heat release model}, which, when coupled with the acoustics in a positive feedback loop, can generate self-excited oscillations and rich nonlinear behaviours via bifurcations \citep[e.g.,][]{Rayleigh1878, dowling2005feedback,  juniper2018sensitivity, Magri2019, magri_linear_2023}. Because these oscillations can have detrimental effects on the system's structure and performance, their prediction and control are active areas of research, for example, in gas turbines \citep{Lieuwen2006,Poinsot2017}, and rocket engines \citep{Culick2006}.
Traditionally, the prediction of thermoacoustics has been achieved with first principles. 
In the time domain, a direct approach 
is the brute-force time-integration of the governing equations following a discretization scheme. High-fidelity models such as large-eddy simulations that model the acoustics and the flame simultaneously on fine grids provide highly accurate solutions, but they are computationally expensive \citep{Poinsot2017}. Low-fidelity models reduce the computational effort at the expense of accuracy while obtaining models that can be used for stability, bifurcation analysis, and parametric studies. Generally, these approaches combine a linear acoustic solver with a \rev{heat release model}. In this direction, nonlinear behaviour of longitudinal and annular combustors has been investigated by employing network models based on the travelling-wave ~\citep[e.g.,][]{Dowling1999, Dowling2003, Li2015, bauerheim_analytical_2014, orchini_effects_2019}, Galerkin decomposition of pressure and velocity using only a finite number of acoustic eigenmodes and projection of the PDEs onto those modes ~\citep[e.g.,][]{Zinn1971, Balasubramanian2008}, and numerical discretizations of the PDEs \citep{Sayadi2014,Huhn2020}. Predictions with first principles only are either computationally expensive (e.g., large-eddy simulation), or as good as the model assumptions. Because thermoacoustics is a multi-physics phenomenon, the model assumptions made in low-order models are inevitably substantial. This motivates adding data into the first-principles modelling of thermoacoustics, for which machine learning methods excel. The data typically comes from laboratory experiments that are conducted with setups consisting of a duct with a heat source, which can be a flame or an electrically heated wire mesh~\citep[e.g.,][]{Kabiraj2012,matveev2003thermoacoustic}. In the experiments, the collected data is usually the acoustic pressure measured by microphones at high sampling rates.
In this direction, data assimilation techniques have been applied to improve physics-based qualitatively accurate models of a prototypical thermoacoustic system, whilst estimating model parameters. \citet{Novoa2021} combined a thermoacoustic low-order model with an ensemble Kalman filter, and an Echo State network for real-time bias-aware data assimilation of the acoustic state~\citep{novoa2023bias}. In other applications, neural networks have been developed for \rev{heat release model} or flame response inference in thermoacoustic applications \citep{Selimefendigil2011, Jaensch2017, Tathawadekar2021}. In \citep{Ozan2023}, a physics-informed feedforward neural network approach was developed for learning thermoacoustic limit cycles by using periodic activation functions. \rev{Physics-informed neural networks were also shown to be successful for the reconstruction of acoustic fields with  non-ideal boundary conditions \citep{silvagarzon2023ReconstructionAcousticFields}}.

The overarching goal of this paper is to generalize and develop acoustic neural networks to embed the structure of the nonlinear wave solution into the architecture. 
The specific objective of this paper is three-fold: to (i) predict and extrapolate in time thermoacoustic oscillations, (ii) reconstruct pressure and velocity over the entire domain from pressure sensors only, and (iii) obtain a model that is robust to noise and generalizable to unseen scenarios. 
%
%
%
This paper is organized as follows. In Section \ref{sec:background}, we provide the  mathematical background for feedforward neural networks and Galerkin decompositions. In Section \ref{sec:base_galnn}, we introduce Galerkin neural networks. In Section \ref{sec:application_acoustics}, we discuss the application of Galerkin neural networks to acoustics and thermoacoustics. In Sections \ref{sec:results_rijke} and \ref{sec:results_higher}, we show results for twin experiments on synthetic data from a Rijke tube and on synthetic data from a higher-fidelity model. The paper ends with a conclusion section.

\section{Background}\label{sec:background}
\subsection{Standard feedforward neural networks}\label{sec:FNN}
\begin{figure}[t!]
\centering
\includegraphics[width = 0.5\linewidth]{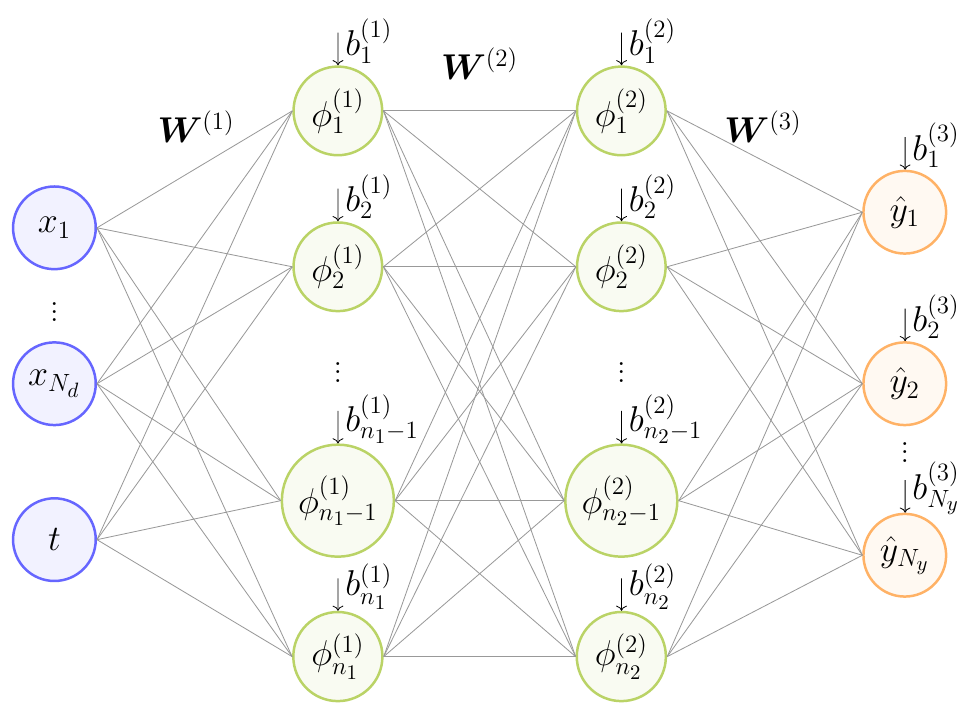}
\caption{Example of a standard feedforward neural network (FNN) ~\citep[e.g.,][]{magri_luca_2023_7655872} with two hidden layers. This is the data-driven only model, which has no prior knowledge embedded in the architecture. The trainable parameters are the weights, $\bm{W}^{(.)}$, and the biases, $\bm{b}^{(.)}$.}  
\label{fig:fnn_architecture}
\end{figure}
%
Let $\bm{y} \in \mathbb{R}^{N_y}$ represent some vector of physical quantities pertaining to a system that depend on space, $\bm{x} \in \mathbb{R}^{N_d}$, where $N_d$ is the dimension, and time, $t \in \mathbb{R}$. Then, given full or partial observations of $\bm{y}$, our goal is to learn a model $\bm{f}$ that predicts an output vector $\bm{\hat{y}}\in \mathbb{R}^{N_y} $ from an an input vector $(\bm{x}, t)$ while minimizing an error metric, $\mathcal{L}$.
A feedforward neural network (FNN) is defined by a composition of functions, $\bm{f}$, which, when appropriately designed, can approximate any continuous function in a specified range~\citep{Hornik1989} 
\begin{subequations}
    \begin{equation}
        \bm{f}(\bm{x}, t) := \bm{f}^{(L)}\left(\bm{f}^{(L-1)}\left(\hdots \bm{f}^{(1)}\left(\bm{x}, t\right)\right)\right), \label{eq:fer3r21}
    \end{equation}
    \begin{equation}
        {\bm{f}}^{(l)}(\bm{z}) := \bm{\phi}^{(l)}\left({\bf W}^{(l)}\bm{z}+\bm{b}^{(l)}\right), \label{eq:fer3r21_1}
    \end{equation}
\end{subequations}
where $ {\bm{f}}^{(l)}$ is the function that maps the layer $l-1$ to the layer $l$, where 
$l=1,\ldots, L$ and $L$ is the number of layers; 
${\bm W}$ are the weights matrices and ${\bm b}$ are the biases, which are the trainable parameters; and  
${\bm{\phi}}^{(l)}$ are the activation functions from the layer $l-1$ to the layer $l$, which are applied to each component of the argument. For regression, in the last layer, the activation ${\bm{\phi}}^{(L)}$ is linear. The neural network offers 
an ansatz for a continuous function through linear operations and simple nonlinear activations~(please, refer to \citep{magri_luca_2023_7655872} for a pedagogical and geometric explanation). 

The FNN is a standard architecture that is fully data-driven, i.e., no prior knowledge is embedded in the network (Figure \ref{fig:fnn_architecture}).
%
The network's weights and biases, collectively grouped in a variable $\bm{\chi}$, are optimized via gradient descent to minimize an error, which is provided by a loss function $\mathcal{L}$  
\begin{equation} \label{eq:dd3o233}
   {\bm{\chi^*} = \underset{\bm{\chi}}{\mathrm{arg \, min}}\;\mathcal{L}(\bm{\chi})}.
\end{equation}
When no physics-constraint is imposed as in standard neural networks, the data-driven loss is quantified by the mean-squared error (MSE) between the measured data and predictions of the network. In the formulation of the data-driven loss, we also account for the cases of partial state observations with a measurement matrix $\bm{M} \in \mathbb{R}^{N_y \times N_y}$ that indicates which states are measured,
\begin{equation}
    M_{ij} = \begin{cases}
        1 \quad \text{when $i = j$ and $y_i$ is measured}, \\
        0 \quad \text{otherwise}.
    \end{cases}
\end{equation}
When full state measurements are available, then $\bm{M}$ is the identity matrix, $\bm{I}_{N_y \times N_y}$. The loss is given as
\begin{equation}\label{eq:loss_data}
       \mathcal{L} \equiv \mathcal{L}_{DD} 
        = \frac{1}{N N_m}\sum_{k = 1}^{N}
        ||\bm{M}{\bm y}_k-\bm{M}\bm{\hat{y}}_k||_2^2,
\end{equation}
where subscript $DD$ stands for data-driven, $N_m \leq N_y$ is the number of measured states, the subscript 2 denotes the $\ell_2$ norm, 
and the subscript $k$ denotes the $k$-th element in the dataset of $N$ pairs of input and output vectors.
In Section~\ref{sec:acoustic_nn}, we tailor the neural networks to acoustic problems by specifying the data, input vectors, task, activations functions, nonlinear maps, and loss functions. The prior knowledge will be embedded in different ways in the neural networks. 



\subsection{Separation of variables and Galerkin methods}\label{sec:sep_var_galerkin}
Separation of variables seeks to find solutions to partial differential equations in a special form as the product of functions of the individual independent variables, e.g., time and single space coordinates, which results in an eigenvalue problem \citep{OlverBook}. Linear wave equations and wave solutions can thus be represented in an acoustic eigenbasis, namely Fourier modes, which is complete. On the other hand, the thermoacoustic problem is governed by a wave equation with a nonlinearity arising from the heat release (please refer to Section \ref{sec:model} for details). However, the nonlinearity will only slightly change the frequency and mode coupling, which motivates the use of the acoustic eigenspace as an expressive basis to represent the nonlinear solutions as well \citep{Zinn1971}. A weak solution to the nonlinear problem is provided by the Galerkin method, which approximates the solution to the PDE by projecting the equations onto a finite-dimensional subspace \citep{HolmesGalerkin}. The solution is written as a linear combination of the basis functions that span this subspace \citep{HolmesGalerkin},
\begin{equation}\label{eq:galerkin_decomp}
    y_i(x,t) = \sum_{j=1}^{N_g}\alpha_j^{i}(t)\Psi_j^{i}(x), 
\end{equation}
where $\bm{y} \in \mathbb{R}^{N_y}$ is a vector of physical quantities (Section \ref{sec:FNN})  and  $i = 1,2,\dots, N_y$ denotes the $i$-th element of $\bm{y}$, $\Psi_j^{i}(x)$ are the basis functions, or Galerkin modes, $N_g$ is the number of Galerkin modes, and $\alpha_j^{i}(t)$ are the Galerkin amplitudes. The orthogonality of the Galerkin modes guarantees that the approximation error is orthogonal to the space spanned by the finite number of modes retained. This property makes the Galerkin method a suitable choice for the modelling of acoustic problems because the low-frequency modes contain most of the energy and thus, the truncation error can be minimal.

\section{Galerkin neural networks}\label{sec:base_galnn}
\begin{figure}[t!]
\centering
\includegraphics[width = 0.7\linewidth]{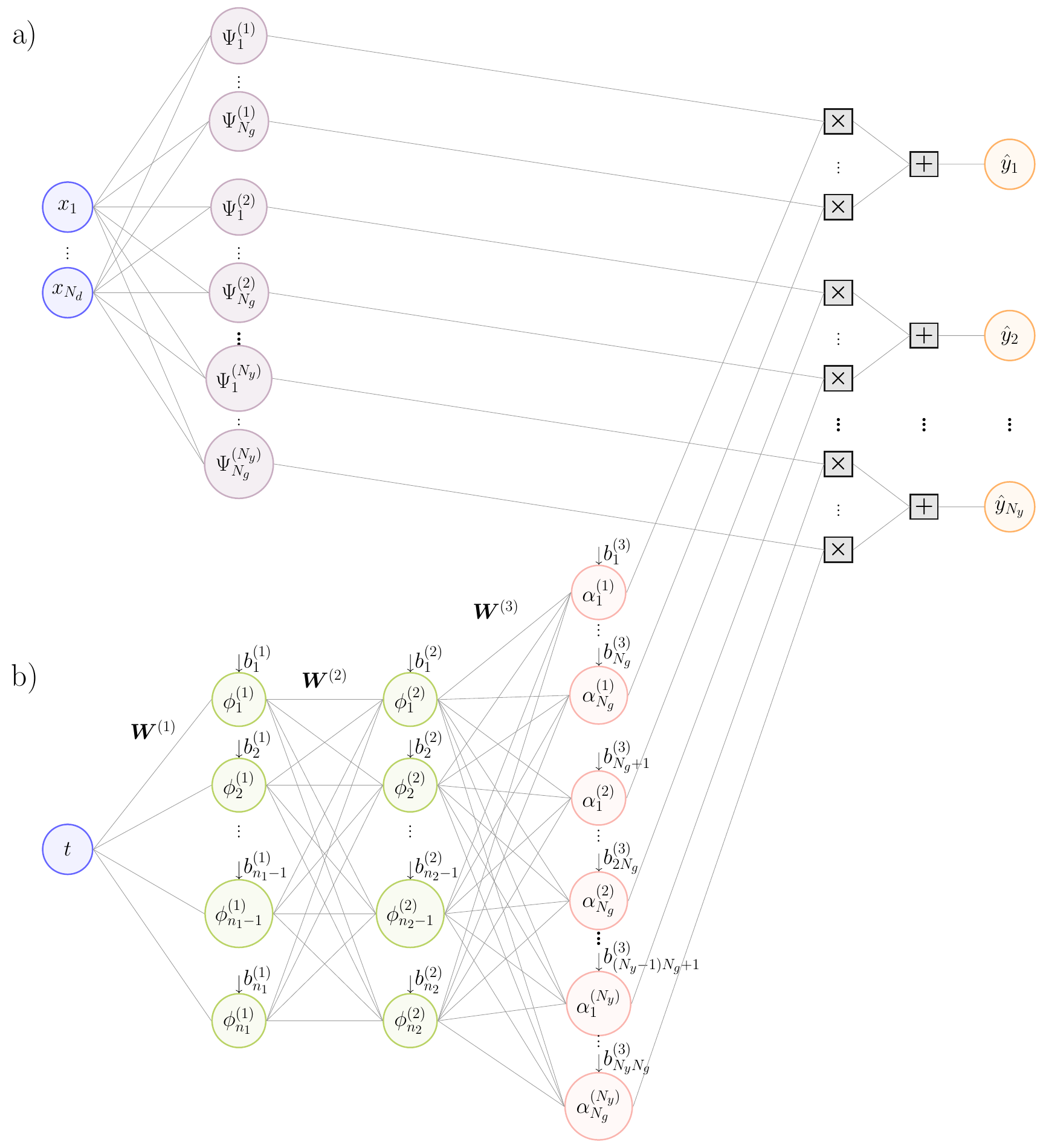}
\caption{Galerkin neural network (GalNN). This network is composed of two branches; a) spatial branch (hard constraint), and b) a temporal branch, which learns the temporal behaviour. The trainable parameters are the weights, $\bm{W}^{(.)}$, and the biases, $\bm{b}^{(.)}$}  
\label{fig:galnn_architecture}
\end{figure}
FNNs are flexible and useful tools for function approximation, however, they can suffer from under- or overfitting, especially in the case of scarce or noisy data. This requires a careful hyperparameter tuning, which becomes computationally expensive in proportion with the size of network. Even then, the network may not be able to capture some features of the solution because of missing data. 
In order to counteract these shortcomings, we exploit the physical knowledge about the spatiotemporal basis of the system in question, and propose a network structure inspired by the Galerkin decomposition of the system as motivated in Section \ref{sec:sep_var_galerkin}. The chosen Galerkin modes are a known nonlinear transformation of the spatial coordinates, which is introduced in the network. Therefore, by design, this network can be configured to automatically satisfy the boundary conditions. We will refer to this network architecture as the \textit{Galerkin neural network} (\textit{GalNN}). The Galerkin network is composed of a spatial branch that transforms $\bm{x}$ into the {\it a-priori} Galerkin modes, $\bm{\Psi}^{i}(\bm{x})$ \eqref{eq:galerkin_decomp}, and a temporal branch that is an FNN \eqref{eq:fer3r21} that takes only $t$ as an input and predicts the unknown Galerkin amplitudes, $\bm{\alpha}^{i}(t)$ \eqref{eq:galerkin_decomp}, as outputs. The final outputs, $\bm{\hat{y}}$, are computed by \eqref{eq:galerkin_decomp}. Formally, the GalNN is defined as
\begin{equation}\label{eq:galnn}
    f_i(\bm{x},t) = \sum_{j=(i-1)N_g+1}^{iN_g} g_j(\bm{x})h_j(t),
\end{equation}
where $\bm{f}$ is the map from $(\bm{x},t)$ to $\bm{\hat{y}}$, $i = 1,2,\dots, N_y$ denotes the $i$-th element of the output vector, $\bm{g}(\bm{x})$ is the spatial branch given by $\bm{g}(\bm{x}) = (\bm{\Psi}^{1}(\bm{x}), \bm{\Psi}^{2}(\bm{x}), \dots, \bm{\Psi}^{N_y}(\bm{x}))$ and $\bm{h}(t)$ is the temporal branch modelled as an FNN, the outputs of which are denoted as $(\bm{\alpha}^{1}(t), \bm{\alpha}^{2}(t), \dots, \bm{\alpha}^{N_y}(t))$. The data-driven loss is defined the same as the standard FNN case \eqref{eq:loss_data} over the prediction error on $\bm{y}$. The time evolution of the Galerkin amplitudes are thus obtained as an intermediary step. This architecture is shown in Figure \ref{fig:galnn_architecture}. 

\section{Application to acoustics and thermoacoustics}\label{sec:application_acoustics}
In this paper, we develop a physics-constrained data-driven model of the acoustic variables as a function of time and space. Given enough capacity by means of number of neurons and layers, FNNs can fit the data within the training range. This can be considered as an interpolation problem, by which we mean that the network predicts on data points that is within the bounds of the training input range. When the prediction is performed for scenarios outside the training input range, we call this an extrapolation problem. Thermoacoustic systems exhibit oscillations, which can be periodic, quasi-periodic, or chaotic in time. First, given time series data from a time window, our task is to obtain a model that can extrapolate such oscillatory behaviour. In experiments, full-state observations may be unavailable or prohibitively expensive, e.g., in acoustics, only pressure data may be available through measurements with microphones. Second, our task is to reconstruct the flow variables over the entire spatial domain from full- or partial noisy state observations, which poses a challenge for purely data-driven models. Third, we seek a robust, generalizable, low-order model for acoustic and thermoacoustic solutions using as little data as possible.

\subsection{Background on thermoacoustics} \label{sec:model}
\begin{figure}[t!]
    \centering
    \includegraphics[width = 0.5\linewidth]{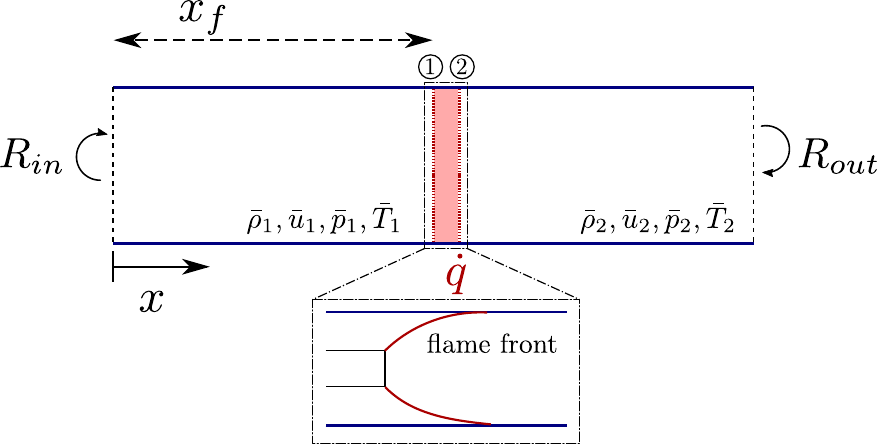}
    \caption{Schematic of the thermoacoustic system. A straight duct with open ends and a heat source, $\dot{q}$, located at $x_f$. The acoustic reflection coefficients at the inlet and outlet are $R_{in}$ and $R_{out}$. The Rijke tube is employed in Section~\ref{sec:results_rijke} for twin experiments with the assumptions; (i) zero Mach number, i.e., $\bar{u}_1=\bar{u}_2=0$, (ii) no change in the mean flow variables before and after the flame e.g., $\bar{\rho}_1=\bar{\rho}_2=1$, and (iii) ideal boundary conditions, i.e., $R_{in}=R_{out}=-1$.  A higher-fidelity model with a mean-flow and a kinematic model for the flame, is employed in Section~\ref{sec:results_higher} for analysing model generalization.}
    \label{fig:rijke}
\end{figure}
We consider a thermoacoustic system composed of a straight duct of length $\tilde{L}$ with a compact heat source located at $\tilde{x} = \tilde{x}_f$, where $\tilde{(\cdot)}$ denotes a dimensional variable (Figure \ref{fig:rijke}). We make the following assumptions about the system; (i) the acoustics are one-dimensional, i.e.,  the tube is sufficiently longer than its diameter for the cut-on frequency to be large enough for longitudinal acoustics only to propagate,  (ii) the heat release from the flame acts as a pointwise monopole source of sound (compact assumption), (iii) the mean-flow is low Mach number and  there is no entropy wave convection, and (iv) effects of viscosity and heat conduction are negligible~(e.g., \cite{Magri2019}). The reflection of the acoustic waves at the boundaries are given by the reflection coefficients \rev{$R_{in}$ and $R_{out}$} for inlet and outlet, respectively, which determine the boundary conditions. 

The dynamics are governed by the dimensional equations derived from mass, momentum, and energy conservation. Modelling the heat source as a compact source results in two duct segments related by jump conditions that are enforced at the heat source location. {For brevity, the suffices $1$ and $2$ denote conditions before and after the flame, i.e., $\tilde{x} = \tilde{x}_{f,1}$ and $\tilde{x} = \tilde{x}_{f,2}$, respectively.} The jump conditions are found from mass, momentum, and energy fluxes across the flame with the ideal gas law \citep[e.g.,][]{Aguilar2019}. The governing equations and the jump conditions are linearized by assuming that the flow variables can be expressed as infinitesimal  perturbations on top of a mean-flow, i.e., $\tilde{(\cdot)} = \tilde{\bar{(\cdot)}} + \tilde{(\cdot)}'$, where $\bar{(\cdot)}$ denotes the steady mean-flow variable and $(\cdot)'$ denotes the unsteady infinitesimal perturbations. Under the low Mach number assumption, the acoustics are governed by momentum and energy equations \citep[e.g.,][]{Magri2019}
\begin{subequations}\label{eq:dim_pde_prime}
    \begin{equation}\label{eq:dim_momentum_prime}
    \tilde{\bar{\rho}}\dtUTildePrime+\tilde{\bar{\rho}}\tilde{\bar{u}}\dxUTildePrime + \dxPTildePrime = 0, 
    \end{equation}
    \begin{equation}\label{eq:dim_energy_prime}
    \dtPTildePrime +\tilde{\bar{u}}\dxPTildePrime + \gamma\tilde{\bar{p}}\dxUTildePrime - (\gamma-1)\frac{\tilde{\dot{q}}'}{\tilde{A}}\delta(\tilde{x}-\tilde{x}_f) = 0, 
    \end{equation}
\end{subequations}
where $\tilde{\rho}$ is the density; $\tilde{u}$ is the velocity; $\tilde{p}$ is the pressure; $\tilde{\dot{q}}$ is the heat release rate; $\tilde{A}$ is the cross-sectional area of the duct; $\gamma$ is the heat capacity ratio; and $\delta$ is the Dirac delta distribution. 
In Sections \ref{sec:results_rijke} and \ref{sec:results_higher}, the governing equations are employed with different simplifications to generate the synthetic data for our study. We will impose the governing equations in their non-dimensional PDE form, denoted by omitting the symbol $\tilde{(.)}$, as soft constraints during the training of our neural networks. This approach will be discussed in detail in Section \ref{sec:pi_loss}. 

\subsection{Acoustic neural networks}\label{sec:acoustic_nn}
The dynamics of thermoacoustic oscillations are dominated by unstable eigenfunctions~\citep[e.g.,][]{Magri2019}, which are periodic oscillations in time. 
We propose neural networks that can naturally infer acoustic dynamics, which are the basis functions of nonlinear thermoacoustic behaviours. 
The proposed networks embed the prior knowledge through the activation functions (Section~\ref{sec:activation}), through a penalization loss function in the training (soft constraint, Section~\ref{sec:pi_loss}), and through the architecture (hard constraint, Section~\ref{sec:galnn}).
For the thermoacoustic system described in \ref{sec:model}, the input vector consists of the one-dimensional spatial coordinate, $x$, and time, $t$, and the output vector consists of the acoustic pressure and velocity fluctuations, $\bm{y} = (p',u')$. In Section~\ref{sec:activation}, we motivate employing periodic activation functions in the standard FNN for acoustic problems, the eigenfunctions of which are periodic. Further, in Section~\ref{sec:pi_loss}, we include a physics-based regularization term that penalizes solutions that violate the conservation laws governing the acoustic dynamics. Finally, in Section~\ref{sec:galnn}, we promote a hard-constrained architecture in the form of a GalNN (Section~\ref{sec:base_galnn}). Inspired by the physical remarks of Section \ref{sec:sep_var_galerkin}, we design a neural network architecture that spans the Hilbert space with the acoustic eigenfunctions as the Galerkin modes, whilst having trainable parameters for inference and closure of the unknown physical terms.
The acoustic neural networks are summarized at the end of this section in Table \ref{tab:nns}.

\subsubsection{Periodically activated feedforward neural networks (P-FNNs)}
\label{sec:activation}
In order to augment the extrapolation capability, observe that nonlinear thermoacoustic dynamics originate from the nonlinear coupling of acoustic eigenfunctions ~\citep[e.g.,][]{Magri2019}, which are Fourier modes. From a data-driven perspective, this means that the weights and biases should be periodically activated. 
A straightforward strategy is to employ periodic activations in the acoustic neural networks, i.e., ${\bm \phi}^{(l)} = \sin({\bm z})$ in Eq.~\ref{eq:fer3r21_1} \citep{Ozan2023}. 
The physics of the system, namely the periodic nature of the solutions, is embedded in the network itself via the choice of activation function, which provides an inductive bias on the function space of the network and improves extrapolation \citep{Ziyin2020}.
%
The weights of a layer $l$ with the sine activation are initialized from a uniform distribution in the range $[-\sqrt{{3}/{\mathrm{fan}_{\mathrm{in}}}},\sqrt{{3}/{\mathrm{fan}_{\mathrm{in}}}}]$, where $\mathrm{fan}_{\mathrm{in}}$ is the number of neurons in the layer $l-1$, $n_{l-1}$~\citep{Ziyin2020}. 

\subsubsection{Soft constraints with physics-informed losses}\label{sec:pi_loss}
Prior knowledge can be embedded as a penalization term in the loss function in Eq.~\ref{eq:dd3o233} to minimize during training~\citep[e.g.,][]{lagaris1998artificial,Raissi2019}. This approach improves the generalizability of the network, while promoting physical outputs \citep{Karniadakis2021}. In thermoacoustics, neural network predictions should fulfill the acoustic conservation laws, therefore, we penalize solutions that violate momentum and energy equations in the loss function
\begin{equation}
    \mathcal{L} = \lambda_{DD}\mathcal{L}_{DD}+\lambda_M\mathcal{L}_M+\lambda_E\mathcal{L}_E,
\end{equation}
where $\mathcal{L}_{DD}$ is the data-driven loss \eqref{eq:loss_data} (Section~\ref{sec:FNN}), and $\mathcal{L}_M$ and $\mathcal{L}_E$ are the residual losses from the conservation of momentum and energy equations, respectively. The non-negative scalars $\lambda_{DD}$, $\lambda_M$, and $\lambda_E$ are regularization hyperparameters.
Obtaining momentum and energy losses, $\mathcal{L}_M$ and $\mathcal{L}_E$, requires the evaluation of the physical residuals of the network predictions. We denote the partial differential operators that define the residual from momentum and energy equations as $\mathcal{F}_M(p',u')$ and  $\mathcal{F}_E(p',u')$, respectively. A given $(p',u')$ is a solution of the system if the residuals are zero, i.e., $\mathcal{F}_M(p',u') = 0$ and $\mathcal{F}_E(p',u') = 0$. We will exactly define these residuals separately for the Rijke tube in Section \ref{sec:rijke_tube} and for the higher-fidelity model in \ref{sec:higher_model} along with the modelling assumptions and non-dimensionalization of the original equations \eqref{eq:dim_pde_prime}. The physics-informed losses are then computed as
\begin{align}
        \mathcal{L}_{M} &= \frac{1}{N+N_s}\sum_{k = 1}^{N+N_s}\mathcal{F}_M(\hat{p}'_k,\hat{u}'_k)^2, \label{eq:loss_momentum} \\ 
        \mathcal{L}_{E} &= \frac{1}{N+N_s}\sum_{k = 1}^{N+N_s}\mathcal{F}_E(\hat{p}'_k,\hat{u}'_k)^2. \label{eq:loss_energy}
\end{align}
where $N$ is the number of training data points, i.e., $\{x_k,t_k\}_{k = 1}^{N}$, and $N_s$ is the number of uniformly sampled points over the whole training domain, as we can evaluate the physical loss at any location in time and space. For this purpose, we employ automatic differentiation using $\texttt{tf.GradientTape}$ functionality from TensorFlow \citep{tensorflow2015-whitepaper}. The automatic differentiation of the output variables with respect to the input variables yields the Jacobian that contains the partial derivatives, $\frac{\partial \hat{p}'}{\partial x}, \frac{\partial \hat{p}'}{\partial t}, \frac{\partial \hat{u}'}{\partial x}, \frac{\partial \hat{u}'}{\partial t}$. These partial derivatives are plugged in the operators, $\mathcal{F}_M$ and $\mathcal{F}_E$, to calculate the residual of the network. 

\subsubsection{Hard-constraints with Galerkin neural networks \it{(GalNNs)}}\label{sec:galnn}
A suitable eigenbasis for the thermoacoustic problem is provided by the natural acoustic eigenfunctions, because the most of the energy of the solution is dominated by the first acoustic modes, and effects of mean flow or nonlinearity only slightly change the frequency and the eigenshapes (Section \ref{sec:sep_var_galerkin}). The core eigenfunctions of the thermoacoustic system with zero Mach number case are similar to a low Mach number case, $M \lesssim 0.1$ \citep{Magri2014}. Hence, to derive them, we assume a zero Mach number for simplification, whose dynamics are governed by the non-dimensional equations \citep[e.g.,][]{Magri2014},
\begin{subequations}\label{eq:rijke_pde}
    \begin{equation}\label{eq:rijke_momentum}
    \bar{\rho}\frac{\partial u'}{\partial t} + \frac{\partial p'}{\partial x} = 0,  
    \end{equation}
    \begin{equation}\label{eq:rijke_energy}
    \frac{\partial p'}{\partial t} + \frac{\partial u'}{\partial x} - \dot{q}' \delta(x-x_f) = 0, 
    \end{equation}
\end{subequations}
where the density is modelled as 
    $\bar{\rho} = \bar{\rho}_1$ when $0\leq x < x_f$, and  $\bar{\rho} = \bar{\rho}_2$ when $x_f <x \leq 1$. 
The dimensional variables have been scaled as, $x = \tilde{x}/\tilde{L}$; $t = \tilde{t}\tilde{\bar{c}}_{ref}/\tilde{L}$, where $\tilde{\bar{c}}_{ref}$ is the mean speed of sound;  $u = \tilde{u}/\tilde{\bar{c}}_{ref}$; $\rho = \tilde{\rho}/\tilde{\bar{\rho}}_{ref}$; $p = \tilde{p}/(\tilde{\bar{\rho}}_{ref}\tilde{\bar{c}}_{ref}^2)$; $\dot{q} = \tilde{\dot{q}}(\gamma-1)/(\tilde{\bar{\rho}}_{ref}\tilde{\bar{c}}_{ref}^3)$.  
A physically-justified method to solve the set of PDEs \eqref{eq:rijke_pde} is to decompose pressure and velocity as
\begin{subequations}\label{eq:rijke_galerkin}
    \begin{equation}\label{eq:rijke_galerkin_p}
        p'(x,t) = \sum_{j=1}^{N_g}\begin{cases}
        \mu_j(t)\Pi_j^{(1)}(x), \quad 0\leq x < x_f, \\
        \mu_j(t)\Pi_j^{(2)}(x), \quad x_f < x \leq 1, 
        \end{cases}
    \end{equation}
    \begin{equation}\label{eq:rijke_galerkin_u}
        u'(x,t) = \sum_{j=1}^{N_g}\begin{cases}
        \eta_j(t)\Upsilon_j^{(1)}(x), \quad 0\leq x < x_f, \\
        \eta_j(t) \Upsilon_j^{(2)}(x), \quad x_f < x \leq 1, 
        \end{cases}
    \end{equation}
\end{subequations}
and project the equations onto the acoustic eigenfunctions found as below in the case of ideal boundary conditions, i.e., $p'(x = 0) = p'(x = 1) = 0$., \cite{Magri2014}
\begin{subequations}\label{eq:rijke_galerkin_modes}
    \begin{align} 
        \Pi_j^{(1)}(x) &= -\sin(\omega_j\sqrt{\bar{\rho}_1}x),  \quad
        \Pi_j^{(2)}(x) =-\left(\frac{\sin\gamma_j}{\sin\beta_j}\right)\sin(\omega_j\sqrt{\bar{\rho}_2}(1-x)), \label{eq:pi_mode} \\ 
        \Upsilon_j^{(1)}(x) &= \frac{1}{\sqrt{\bar{\rho}_1}}\cos(\omega_j\sqrt{\bar{\rho}_1}x), \quad
        \Upsilon_j^{(2)}(x) = -\frac{1}{\sqrt{\bar{\rho}_2}}\left(\frac{\sin\gamma_j}{\sin\beta_j}\right)\cos(\omega_j\sqrt{\bar{\rho}_2}(1-x)), 
        \label{eq:upsilon_mode}
    \end{align}
\end{subequations}
where $N_g$ is the number of Galerkin modes retained in the approximation, and
$\gamma_j = \omega_j\sqrt{\bar{\rho}_1}x_f, \quad \beta_j = \omega_j\sqrt{\bar{\rho}_2}(1-x_f).$
The acoustic angular frequencies $\omega_j$ are the solutions of the dispersion relationship, 
    $\sin\beta_j\cos\gamma_j + \cos\beta_j\sin\gamma_j\sqrt{\frac{\bar{\rho}_1}{\bar{\rho}_2}} = 0.$
In the limit of no jump in mean-flow density, i.e., $\bar{\rho}_1 =\bar{\rho}_2$, $\omega_j = j\pi$, which are the natural acoustic angular frequencies \citep{Magri2014}.  

\begin{figure}[t!]
\centering
\includegraphics[width = \linewidth]{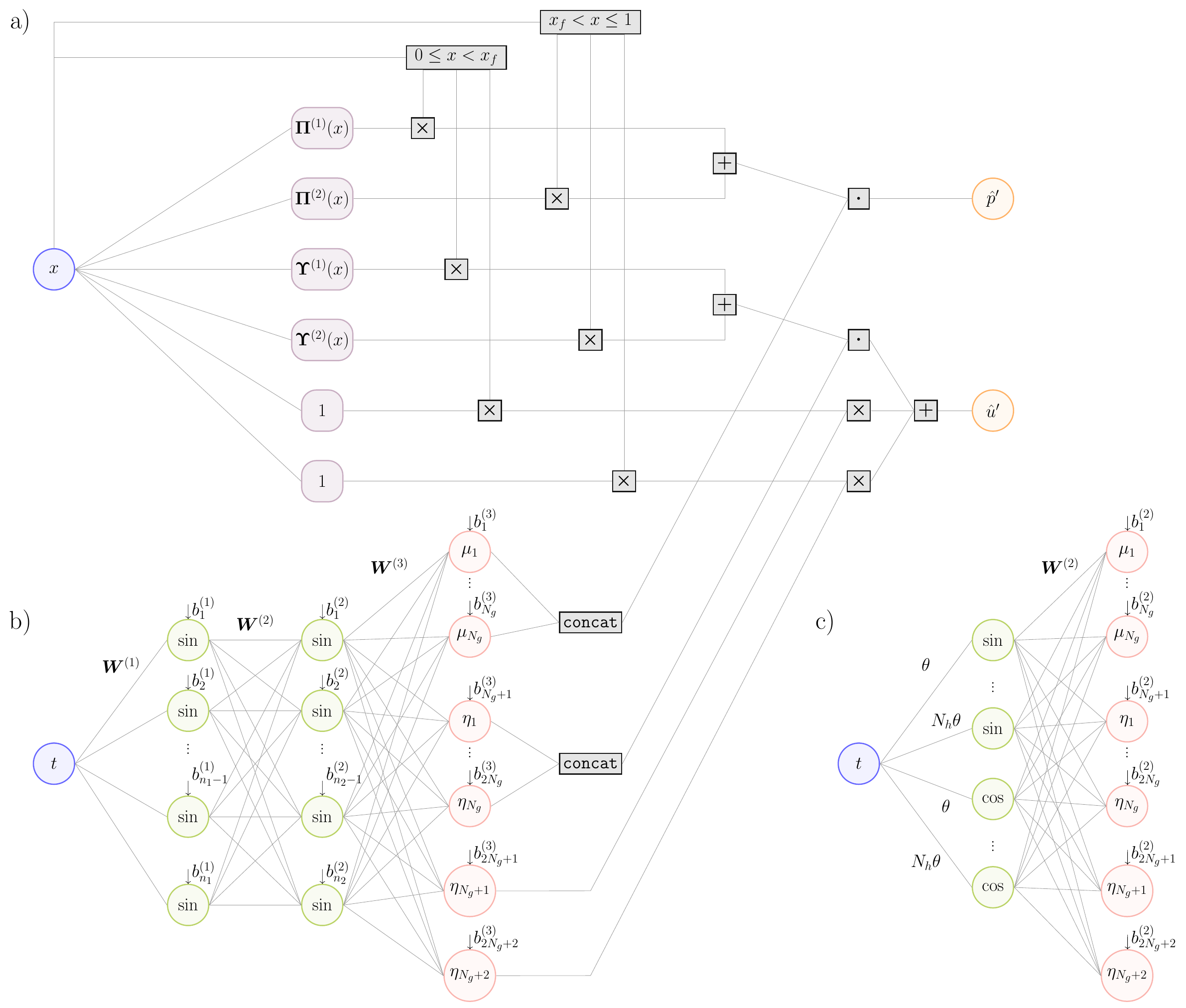}
\caption{The Galerkin neural network (GalNN) tailored for thermoacoustics. (a) Spatial branch, in which the solution space is spanned by the acoustic eigenfunctions \eqref{eq:rijke_galerkin_modes} (hard constraint). The spatial modes $\Pi_j^{(1)}$, $\Pi_j^{(2)}$, $\Upsilon_j^{(1)}$, $\Upsilon_j^{(2)}$, where $j = 1,2,..,N_g$, are collected in the vectors $\bm{\Pi}^{(1)}$, $\bm{\Pi}^{(2)}$, $\bm{\Upsilon}^{(1)}$ and $\bm{\Upsilon}^{(2)}$, respectively. Step functions that model the discontinuity through the compact heat source \eqref{eq:discontinuity_modes} are shown at the bottom of the spatial branch.
(b) Temporal branch, which is a periodically activated feedforward neural network. The outputs $\mu_j$ and $\eta_j$ are stacked as vectors, which are projected onto the modes $\bm{\Pi}^{(1)}$, $\bm{\Pi}^{(2)}$, $\bm{\Upsilon}^{(1)}$ and $\bm{\Upsilon}^{(2)}$. The outputs $\eta_{N_g+1}$ and $\eta_{N_g+2}$ are associated with the discontinuity modes \eqref{eq:discontinuity_modes}. The trainable parameters are the weights, $\bm{W}^{(\cdot)}$, and the biases, $\bm{b}^{(\cdot)}$
(c) Temporal branch of the periodic Galerkin neural network further constrained for limit-cycle solutions. It is reconstructed such that the weights are integer multiples of only one trainable fundamental angular frequency, $\theta$. The trainable parameters are $\theta$, the weights, $\bm{W}^{(2)}$, and the biases, $\bm{b}^{(2)}$}
\label{fig:galerkin_nn_architecture}
\end{figure}
Motivated by the Galerkin decomposition \eqref{eq:rijke_galerkin}, we express the acoustic and thermoacoustic solutions as a GalNN \eqref{eq:galnn}, where $\Pi_j^{(1)}$, $\Pi_j^{(2)}$ \eqref{eq:pi_mode}, and $\Upsilon_j^{(1)}$, $\Upsilon_j^{(2)}$ \eqref{eq:upsilon_mode} are the Galerkin modes, and $\mu_j$ and $\eta_j$ are the Galerkin amplitudes.
The pressure and velocity are determined from the predicted Galerkin amplitudes using \eqref{eq:rijke_galerkin}. This architecture is shown in Figure \ref{fig:galerkin_nn_architecture} (a, b). 

Whereas the acoustic pressure is continuous across the flame, the acoustic velocity undergoes a discontinuity due to the dilation from the heat-release rate \citep{Magri2019}. 
When such a discontinuity is approximated with a finite Fourier series, such as the Galerkin modes we are using, the Gibbs phenomenon manifests itself as high-frequency oscillations around the discontinuity \citep{Magri2013,Sayadi2014}, which causes an unphysical behaviour.
To physically capture the discontinuity and eliminate unphysical oscillations in the predictions, we add two step functions in the velocity modes
\begin{subequations}\label{eq:discontinuity_modes}
    \begin{align}
        \Upsilon_{N_g+1}^{(1)}(x) &= 1, \quad \Upsilon_{N_g+1}^{(2)}(x) = 0, \\
        \Upsilon_{N_g+2}^{(1)}(x) &= 0, \quad \Upsilon_{N_g+2}^{(2)}(x) = 1,
    \end{align}
\end{subequations}
such that the summation \eqref{eq:rijke_galerkin_u} runs from 1 to $N_g+2$. The modes $\Upsilon_{N_g+1}^{(.)}$ and $\Upsilon_{N_g+2}^{(.)}$ are weighted by independent coefficients, $\eta_{N_g+1}$ and $\eta_{N_g+2}$, which allows them to capture a jump discontinuity at the flame location. 

The temporal branch of the GalNN should be capable of extrapolation in time of thermoacoustic oscillations. These solutions bifurcate from limit-cycle oscillations, which are periodic, to quasiperiodic and chaotic oscillations. As they originate from periodic solutions, even the quasiperiodic and chaotic solutions are in fact dominated by periodicity, which can also be observed in the frequency spectrum of the oscillations (later in Section \ref{sec:long_term}). Therefore, we deploy a periodically activated FNN (Section~\ref{sec:activation}) as the temporal branch of the GalNN.

\subsubsection{Periodic Galerkin networks \it{(P-GalNNs)}}\label{sec:periodic_galnn}
We enforce periodic activations (Section~\ref{sec:activation}) in the Galerkin neural network in order to achieve extrapolation for limit-cycle, quasiperiodic, and chaotic solutions. For simplicity, if we consider a single-layer periodically activated GalNN, then the outputs of the temporal branch of the network are given as a sum of sinusoids, where the hidden layer weights represent the angular frequencies and the biases represent the phases, i.e., $\sin(\mathbf{{W}}^{(1)}t+\bm{b}^{(1)})$. For a signal given as a sum of sinusoids, its angular frequency is the greatest common divisor of the angular frequencies of its components. For a limit-cycle, the greatest common divisor is its fundamental angular frequency, and the other frequencies are its harmonics. This applies to a periodically activated neural network as well. However, the weights of the neural network are initialised randomly and then trained on a finite amount of data, so it is numerically unlikely for the learned weights to be exact integer multiples of a fundamental frequency. To overcome this numerical challenge, we add a further constraint on the GalNN to guarantee periodic behaviour with the fundamental frequency of the limit-cycle. We create a hidden layer that takes time as input and outputs $(\sin(\bm{h}\theta), \:\cos(\bm{h}\theta)),  \; \bm{h} = (1,2,...,N_h)$, such that the weights in the periodic activations are integer multiples of one trainable variable, $\theta$. The variable $\theta$ is initialized to the non-dimensional angular frequency of the acoustic system, $\pi$, which is an educated guess for the actual thermoacoustic frequency as from our physical knowledge we know that the nonlinearity will only slightly change this. Hence, upon training, $\theta$ corresponds to the angular frequency of the limit-cycle, and the number of harmonics, $N_h$ can be regarded as a hyperparameter. Because we have both sine and cosine activations for the same angular frequency, the phase information is automatically captured, hence we do not require a bias term. The temporal branch of the GalNN tailored for limit-cycles is shown in Figure \ref{fig:galerkin_nn_architecture} (c). 

\begin{table*}
\caption{Summary of the proposed neural networks. All the networks can be equipped with a soft constraint, which penalizes non-physical solutions in the loss function (see Sections \ref{sec:pi_loss}). When the soft constraint is present, the abbreviation will be prefixed with PI (physics-informed). For example, a Galerkin network with the soft constraint will be abbreviated as PI-GalNN.}
\begin{tabular}{llllll}
\hline
Network                                                                                         & Abbreviation & Constraint & \begin{tabular}[c]{@{}l@{}}Separation of \\ time and space\end{tabular}  & Can extrapolate in time & Section \\ \hline
\begin{tabular}[c]{@{}l@{}}Feedforward \\ neural network\end{tabular}                            & FNN & No & No & No & \ref{sec:FNN} \\ \hline
\begin{tabular}[c]{@{}l@{}}Periodically activated \\ feedforward \\ neural network\end{tabular}
                & P-FNN & No & No & Yes - via activation function & \ref{sec:activation} \\ \hline
\begin{tabular}[c]{@{}l@{}}Galerkin \\ neural network\end{tabular}                              & GalNN & Hard & Yes & \begin{tabular}[c]{@{}l@{}}Yes\end{tabular} & \ref{sec:galnn}\\ \hline
\begin{tabular}[c]{@{}l@{}}Periodic Galerkin \\ neural network\end{tabular}                              & P-GalNN & Hard & Yes & \begin{tabular}[c]{@{}l@{}}Yes, for limit-cycles it\\guarantees fundamental frequency\end{tabular} & \ref{sec:periodic_galnn} \\ \hline
\end{tabular}
\label{tab:nns}
\end{table*}

\section{Extrapolation and state reconstruction: Twin experiments}\label{sec:results_rijke}
\subsection{Dataset from the Rijke tube}\label{sec:rijke_tube}
We perform twin experiments on synthetic data generated from a prototypical thermoacoustic system, the Rijke tube, to demonstrate how the acoustic neural networks developed in \ref{sec:acoustic_nn} tackle extrapolation and state reconstruction.
The Rijke tube (Figure~\ref{fig:rijke}) captures the qualitative nonlinear dynamics and bifurcations observed in real-life applications \citep{Balasubramanian2008}.
In this model, we further assume that (i) the mean flow has a zero Mach number with uniform density, and (ii) the boundary conditions are ideal, i.e. $p'(x=0,t)=p'(x=1,t)=0$, hence $R_{in}=R_{out}=-1$. We solve the associated PDEs \eqref{eq:rijke_pde} using the Galerkin decomposition \eqref{eq:rijke_galerkin}-\eqref{eq:rijke_galerkin_modes}, with $\bar{\rho}_1 =\bar{\rho}_2 = 1$, which implies that $\omega_j = j\pi$ \citep{Magri2014}. By substituting the pressure and velocity variables in \eqref{eq:rijke_pde} with their Galerkin decompositions in \eqref{eq:rijke_galerkin} and projecting the dynamics onto the Galerkin modes, the dynamics of the Galerkin variables $\eta_j$ and $\mu_j$ are described by a $2N_g$-dimensional system of ordinary differential equations \citep{Huhn2020}
\begin{subequations}
    \begin{align}
    & \dot{\eta}_j-\mu_j j\pi = 0, \label{eq:eta_dot} \\
    & \dot{\mu}_j+\eta_j j\pi+\zeta_j\mu_j+2\dot{q}'\sin(j\pi x_f) = 0. \label{eq:mu_dot}
    \end{align}
    \label{eq:rijke_galerkin_ode}
\end{subequations}
In the models of Rijke tube in literature \citep[e.g.,][]{Juniper2011}, a modal damping term acting on the pressure, $\zeta p'(x,t)$, is added to the energy equation \eqref{eq:rijke_energy} to account for the unmodelled effects of dissipation at the boundaries, and at viscous and thermal boundary layers at the tube walls. This term lets the higher modes to be damped out. In the projected dynamics, the damping term is $\zeta_j \mu_j$, where $\zeta_j = c_1j^2+c_2j^{1/2}$ \citep{Landau1987}. The heat release rate is described by a modified King's law 
\begin{equation}\label{eq:kings}
    \dot{q}' = \beta\left(\sqrt{|1+u'(x_f,t-\tau)|}-1\right)
\end{equation}
where $\beta$ and $\tau$ are the heat release strength and the flame time delay, respectively \citep{Heckl1990}.
The time-delayed problem is transformed into an initial value problem via an advection function with the dummy variable $v$ so that it can be solved with a standard time-marching scheme \citep{Huhn2020}
\begin{equation}\label{eq:advection}
    \frac{\partial v}{\partial t} + \frac{1}{\tau}\frac{\partial v}{\partial X} = 0, \quad 0\leq X \leq 1, \quad
    v(X = 0,t) = u'(x_f,t).
\end{equation}
The PDE \eqref{eq:advection} is discretized using a Chebyshev spectral method with $N_c$ points \citep{Trefethen2000}. 
\\
The physics-informed loss of the Rijke tube system is calculated by evaluating the left-hand side of the PDEs \eqref{eq:rijke_pde}, i.e., the momentum residual, $\mathcal{F}_M$, is defined as the left-hand side of \eqref{eq:rijke_momentum}, and the energy residual, $\mathcal{F}_E$, is defined as the left-hand side of \eqref{eq:rijke_energy}. When calculating the heat release term \eqref{eq:kings} in the physics-informed loss, the time-delayed velocity at the flame can be predicted from the network as $\hat{u}'(x_f,t_k-\tau)$ (for the GalNN $\hat{u}'(x_f,t_k-\tau) = \sum_{j = 1}^{N_g}\hat{\eta}_j(t_k-\tau)\cos(j\pi x_f)$), i.e., by evaluating the network at $(x = x_f, t = t_k-\tau)$. Therefore, since the model takes time as an input for both FNN and GalNN architectures, there is no need to account for the time delay with a separate dummy variable. The evaluation of the heat-release law requires the system parameters $x_f$, $\beta$, and $\tau$ to be known.

For the FNN, the implementation details concerning the calculation of heat release and modal damping are provided in the Appendix \ref{sec:fnn_pi_appendix}. For the GalNN, as Galerkin amplitudes are available, the physical residuals of the Rijke tube can be directly computed from the ODEs \eqref{eq:rijke_galerkin_ode} in these twin experiments.

\subsection{Extrapolation in time}\label{sec:results_rijke_activation}
\begin{table*}
    \centering
    \caption{Summary of neural network model properties trained on the Rijke tube data.}
    \begin{tabular}{l|l|l|l|l|l|l}
     \hline
         &  ReLU FNN & tanh FNN & sin FNN &  sin-ReLU FNN & GalNN & P-GalNN\\
         \hline
        Hidden layers & 5 & 5 & 3 & 2 & 2 & 1\\
        Activations & ReLU & tanh & sine & sine-ReLU & sine & harmonics \\
        Neurons & 96 & 96 & 32 & 64 & 16 & 40\\
        Learning rate & 0.001 & 0.001 & 0.001 & 0.001 & 0.001 & 0.001\\
        Optimizer & \multicolumn{6}{c}{Adam} \\
        Batch size & \multicolumn{6}{c}{64} \\ \hline
    \end{tabular}
    \label{tab:rijke_nns}
\end{table*}
\begin{figure*}[t!]
    \centering
    \includegraphics[width = \linewidth]{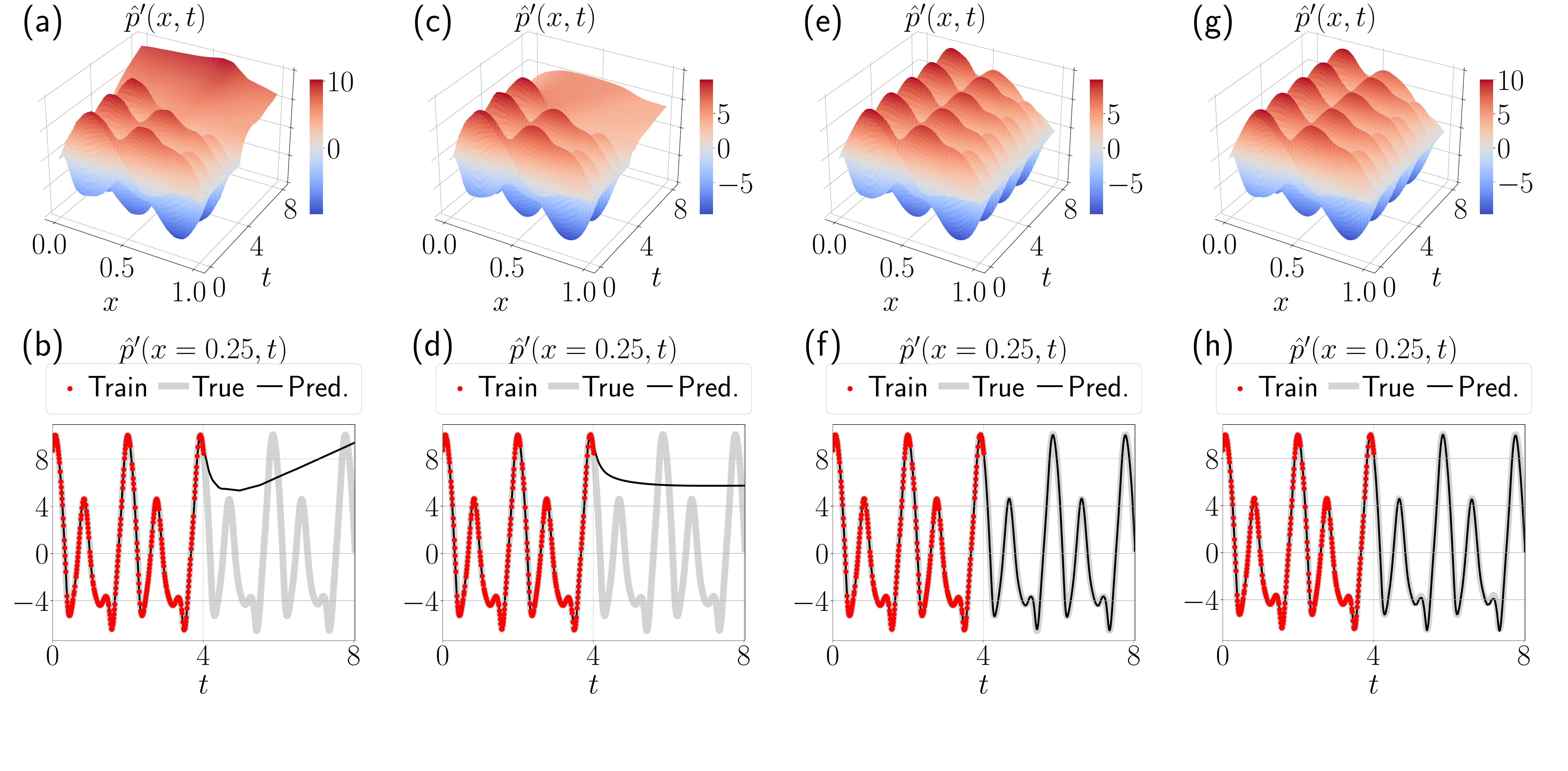}
    \caption{Extrapolation in time of the acoustic pressure with standard neural networks. Effect of activation functions: (a, b) ReLU, (c, d) tanh, (e, f) sine in all hidden layers, (g, h) sine-ReLU (sine in the first layer, ReLU for the rest of the hidden layers). Top row: predictions in both time and space. 
    Bottom row:  predictions in time at $x=0.25$, training data, and ground truth. Typical activation functions, such as ReLU (a, b) and tanh (c, d), fail to extrapolate in time.}
    \label{fig:activation_comparison}
\end{figure*}
We simulate the Rijke tube system \eqref{eq:rijke_pde} with the parameters; $N_g = 10$, $N_c = 10$, $x_f = 0.2$, $\beta = 5.7$, $\tau = 0.2$, $c_1 = 0.1$, $c_2 = 0.06$. After a transient period, this system settles onto a limit-cycle with a period of $\approx1.927$ time units. \rev{The dataset is composed of pressure and velocities measurements from thirteen sensor locations, which are uniformly distributed along the tube including the boundaries.} Before training, the inputs are standardized to have a zero mean and unit variance.  

\subsubsection{Standard feedforward neural networks}
Figures \ref{fig:activation_comparison} (a,b) and (c,d) show the results of FNNs with rectified linear unit (ReLU) and tanh activation functions when trained on this dataset. The network hyperparameters (number of layers, number of neurons, learning rate) have been tuned via a grid search, in which we choose the architectures that resulted in the smallest validation losses. The network properties are provided in Table \ref{tab:rijke_nns}. The ReLU network is initialized with the method of \citet{He2015}, and the tanh network with the method of \citet{Glorot2010}. As illustrated in Figures \ref{fig:activation_comparison} (a,b) and (c,d), FNNs with conventional activation functions fail at extrapolating periodic functions in time.

Figure \ref{fig:activation_comparison} (e,f) shows the extrapolation capability of an FNN equipped with sine activation, i.e., a periodically activated FNN (P-FNN) as introduced in Section \ref{sec:activation}. (Note that when using the sine activation, the inputs are not normalized before training.) The activation function is modified to include a hyperparameter $a$ such that it becomes 
\begin{equation}\label{eq:sine}
    \phi(z) = \frac{1}{a}\sin(az).
\end{equation}
This hyperparameter is used to fine tune the frequency content of the learned functions. For this dataset, it is determined as $a = 10$ with a grid search, the effect of varying $a$ is shown in more detail in the Appendix \ref{sec:hyp_a_appendix}. Physically, since the weights are initialized as $\sim\textit{Uniform}(-1.22,1.22)$ (Section \ref{sec:activation}), multiplying this by $a = 10$, provides a good initial guess on the order of magnitude of the angular frequency for the training. Figure \ref{fig:activation_comparison} (g,h) shows the case for the sine-ReLU configuration, which can also extrapolate in time. \rev{Both sine and sine-ReLU networks can extrapolate also at locations where no sensor measurements are available (result not shown).} Deeper architectures are required when training with purely ReLU or tanh networks compared to sine or sine-ReLU networks. For example, in this case we use two layers of sine and sine-ReLU vs. five layers of ReLU and tanh, and even then the ultimate training and validation losses of sine and sine-ReLU networks were one order of magnitude smaller. To conclude, using periodic activations produces more physical and expressive neural network models for acoustics than the conventional activations, which can also extrapolate in time, whilst speeding up training and hyperparameter search as a consequence of the smaller network sizes required.

\subsubsection{Physics-constrained networks and long-term extrapolation}

\label{sec:long_term}
\begin{figure}[t!]
    \centering
    \includegraphics[width = \linewidth]{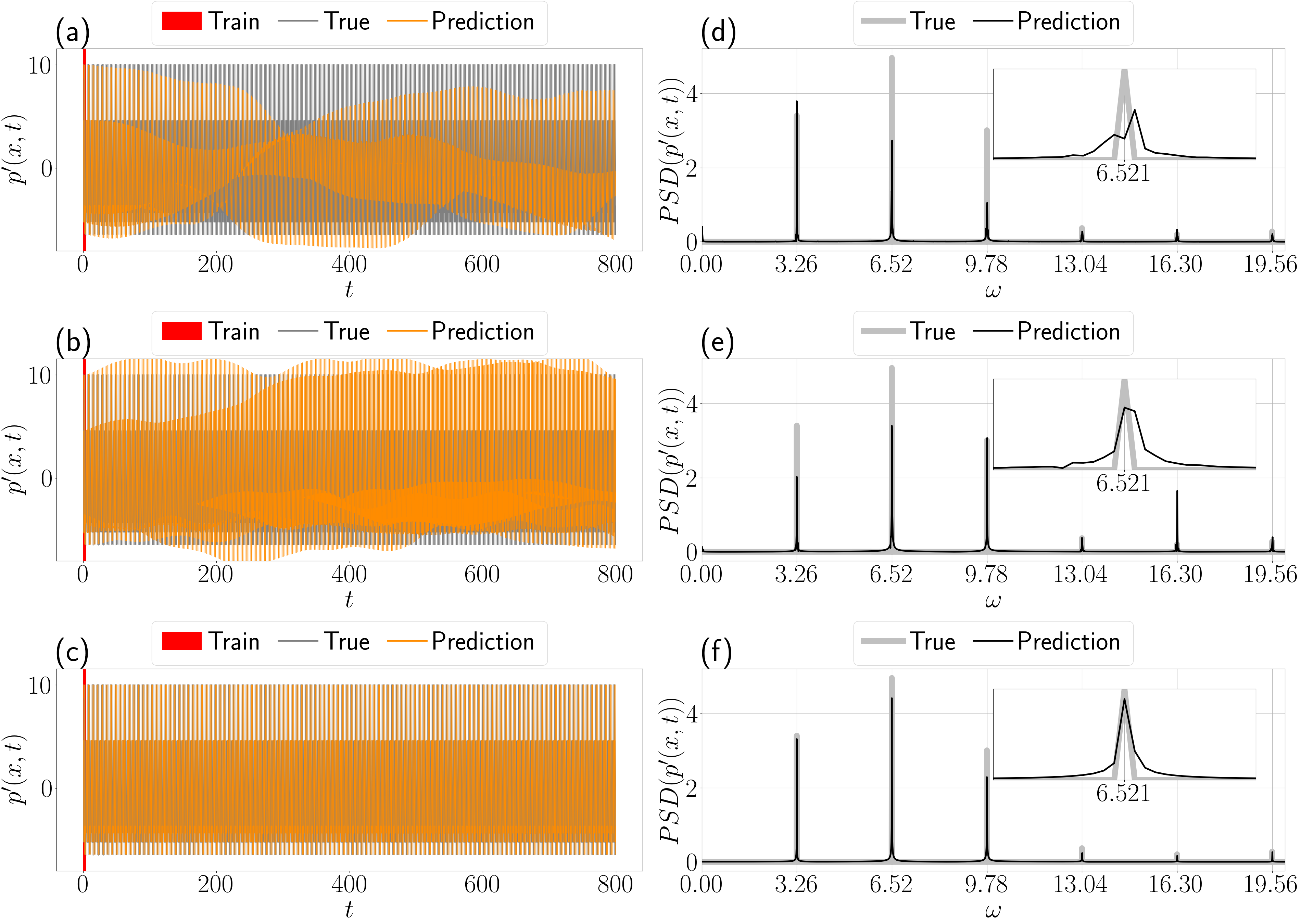}
    \caption{Long-term extrapolation in time of periodic acoustic pressure timeseries with physics constrained neural networks.  The training data is small and spans the first four time units inside the red vertical line. 
    Predictions on the pressure time series with (a) periodically activated feedforward neural network (P-FNN), (b) Galerkin neural network (GalNN), and (c) periodic Galerkin neural network (P-GalNN). Power spectral density of true and predicted time-series with (d) P-FNN, (e) GalNN, and (f) P-GalNN. Constraining prior knowledge on periodicity is key to an accurate long-term prediction.}
\label{fig:long_term}
\end{figure}

\begin{figure}[tbh!]
    \centering
    \includegraphics[width = \linewidth]{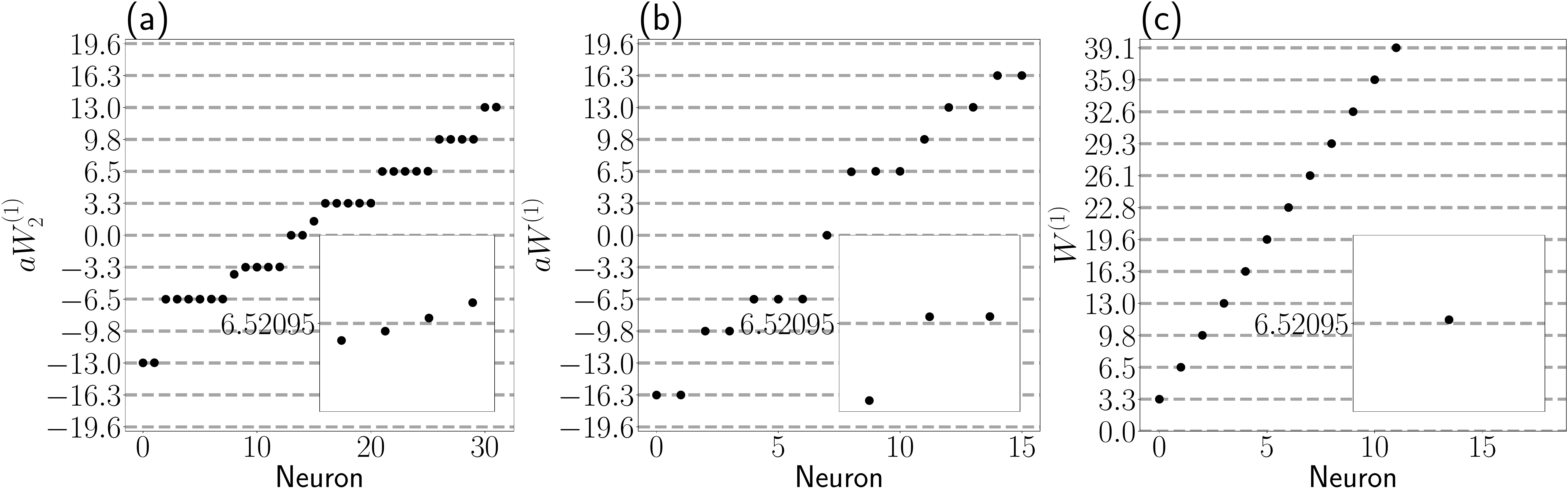}
    \caption{In reference to Fig.~\ref{fig:long_term}, trained weights of the first hidden layers of (a) periodically activated feedforward neural network (P-FNN), (b) Galerkin neural network (GalNN), and (c) periodic Galerkin neural network (P-GalNN). These weights are multiplied with the time input before periodic activation, i.e., $\sin(a\bm{W}^{(1)}_1x+a\bm{W}^{(1)}_2t+a\bm{b}^{(1)})$ for P-FNN, $\sin(a\bm{W}^{(1)}t+a\bm{b}^{(1)})$ for GalNN, where $a$ is a hyperparameter, and $\sin(\bm{W}^{(1)}t)$ for periodic GalNN. }
    \label{fig:weights}
\end{figure}

We showed that networks equipped with periodic activations make accurate predictions on data points chosen in a time range right after the training. Here, we analyse how the learned models extrapolate in the long term. 
The ideal model of a periodic solution should capture the periodicity that the system exhibits for all times. Network predictions over a long time period are plotted in Figure \ref{fig:long_term} (a) and (b), for P-FNN and GalNN, respectively, \rev{when trained on data from approximately two periods}. Over a long time range, we observe a beating-like phenomenon. 
In the frequency spectra of the acoustic variables, the peaks are at the harmonics of the angular frequency of the limit-cycle, which we determined as $\theta^* = 3.2605$ by inspection of the frequency spectrum and autocorrelation of the timeseries. After the training, this is captured by the weights in the first hidden layer as discussed in Section \ref{sec:periodic_galnn}. These weights are multiplied with the time input before periodic activation, i.e., $\sin(a\bm{W}^{(1)}_1x+a\bm{W}^{(1)}_2t+a\bm{b}^{(1)})$ for P-FNN, and $\sin(a\bm{W}^{(1)}t+a\bm{b}^{(1)})$ for GalNN, where $a$ is a hyperparameter of the sine activation \eqref{eq:sine}.
The optimized weights are placed close to the harmonics of the angular frequency of the limit-cycle. Figure \ref{fig:weights} shows the (sorted) weights of the P-FNN and a GalNN. The harmonics of the angular frequency of the limit-cycle are shown in dashed horizontal lines. Although these weights are close to the harmonics of the frequency of the limit-cycle, they are not exact integer multiples of each other. Even the weights clustered around one frequency slightly differ from each other. This can be observed in the power spectral density (PSD) of the predicted timeseries as well, shown in Figure \ref{fig:long_term} (d) and (e). These observations align with the beating behaviour.

We overcome this issue by  
using the GalNN tailored for this specific purpose as described in Section \ref{sec:periodic_galnn}. The long term predictions are shown in Figure \ref{fig:long_term}(c) and the power spectral density of the predicted signal in Figure \ref{fig:long_term}(f). 
The estimated $\theta^*$ by the periodic GalNN is $3.2615$, which aligns with the true signal. The trained weights, which are activated as $(\sin(\bm{W}^{(1)}t, \cos(\bm{W}^{(1)}t)$ in the periodic GalNN, are shown in Figure \ref{fig:weights}(c). \rev{Increasing the training time  improves the extrapolation ability for both P-FNN and GalNN. However, with the constrained structure of the GalNN, we eliminate the beating behaviour and achieve accurate long-term extrapolation without using more training data.}

\begin{figure}[t!]
    \centering
    \includegraphics[width = \linewidth]{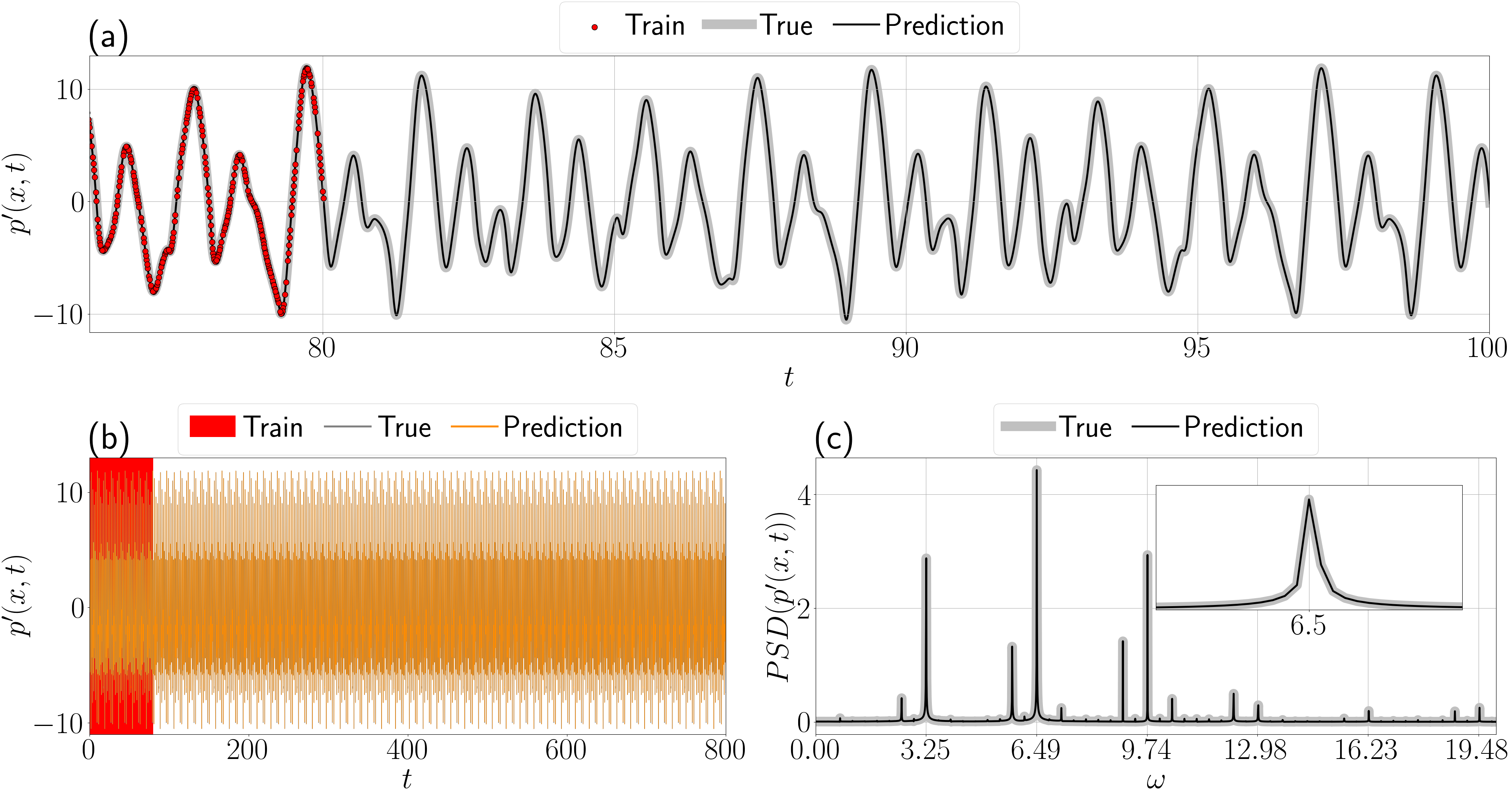}
    \includegraphics[width = \linewidth]{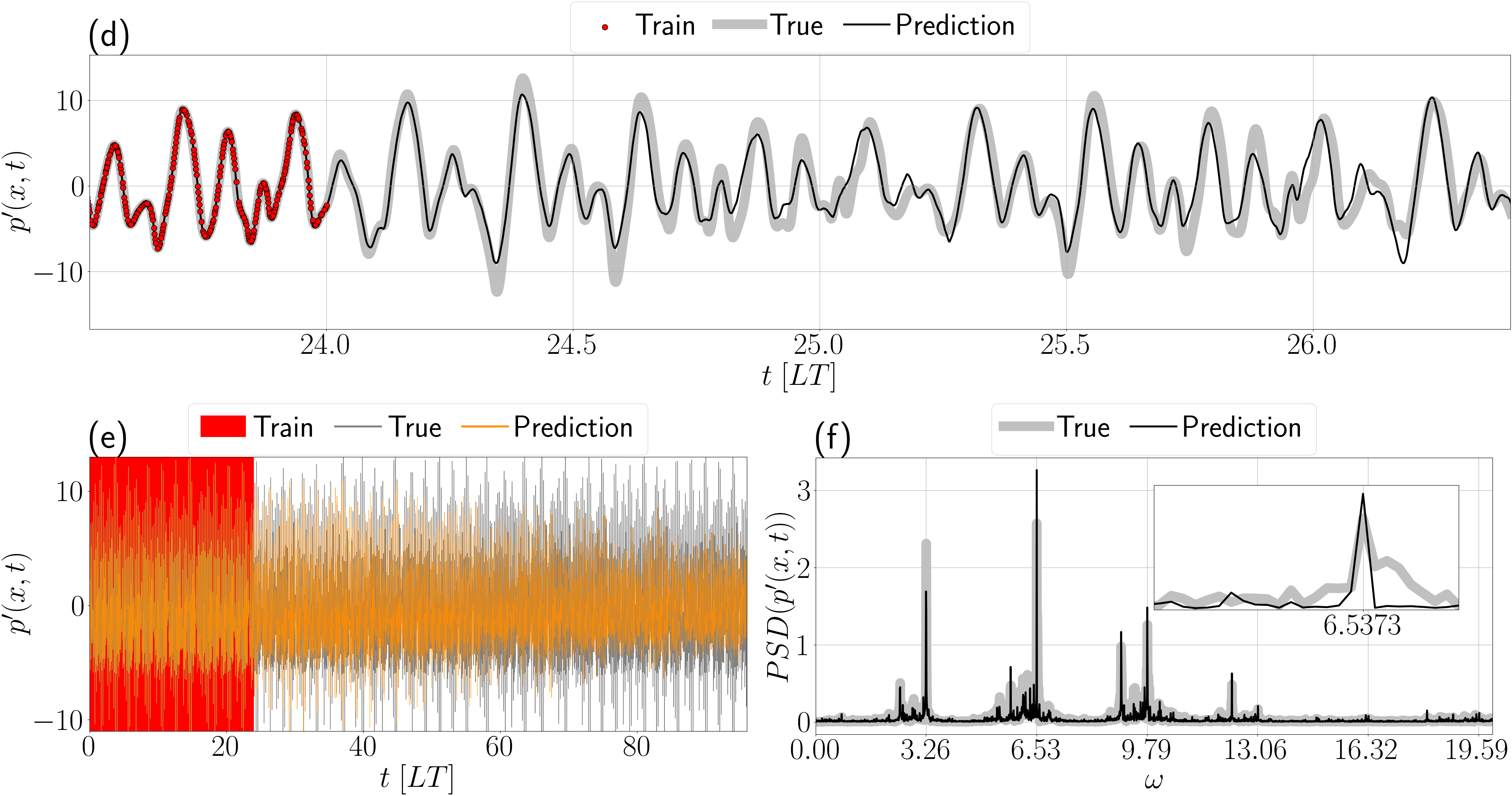}
    \caption{Extrapolation in time for non-periodic solutions. (a-c) Quasiperiodic solution; and (d-f) chaotic solution. \rev{For the chaotic timeseries, the time axis is normalised by the Lyapunov time (LT) of the system, where $\mathrm{LT} \approx 8.33$ time units in this case.} Training data shown within the red vertical line (80 time units for the quasiperiodic case, 200 time units \rev{(24 LT)} for the chaotic case). }
\label{fig:quasi-chaotic}
\end{figure}
In addition to limit-cycles, thermoacoustic systems can also exhibit quasiperiodic and chaotic oscillations through bifurcations \citep[e.g.,][]{Kabiraj2012}. 
With $x_f = 0.2$ and $\tau = 0.2$ and increasing $\beta$, the system takes a Ruelle-Takens-Newhouse route to chaos; at $\beta = 0.5$, it bifurcates from a fixed point solution to a limit-cycle, at $\beta = 5.8$ from a limit-cycle to a quasiperiodic attractor, and at $\beta = 6.5$ from a quasiperiodic attractor to chaos~\citep{Huhn2020}. We collect quasiperiodic data from $\beta = 6$ regime and chaotic data from $\beta = 7$ regime. For the quasiperiodic system, we train a GalNN with 2 sine layers of 96 neurons on data from the 80 time units long time series following a hyperparameter tuning with a grid search. The predictions are shown in Figure \ref{fig:quasi-chaotic}(a-c) with the PSD. The Galerkin network can capture the quasiperiodic behaviour and the frequency spectrum correctly. 
For the chaotic system, we train a GalNN with 4 sine layers of 256 neurons on data from the 200 time units long time series. The predictions are shown in Figure \ref{fig:quasi-chaotic}(d-f) along with the PSD. To train on chaotic data, we required a longer time series and a deeper and wider network because the nonlinear dynamics is richer than the quasiperiodic solution. 
The network's prediction correctly captures the dominant frequencies and the amplitude of the true signal. 

\subsection{Learning from partial state measurements}
\begin{figure}[t!]
    \centering
    \includegraphics[width = \linewidth]{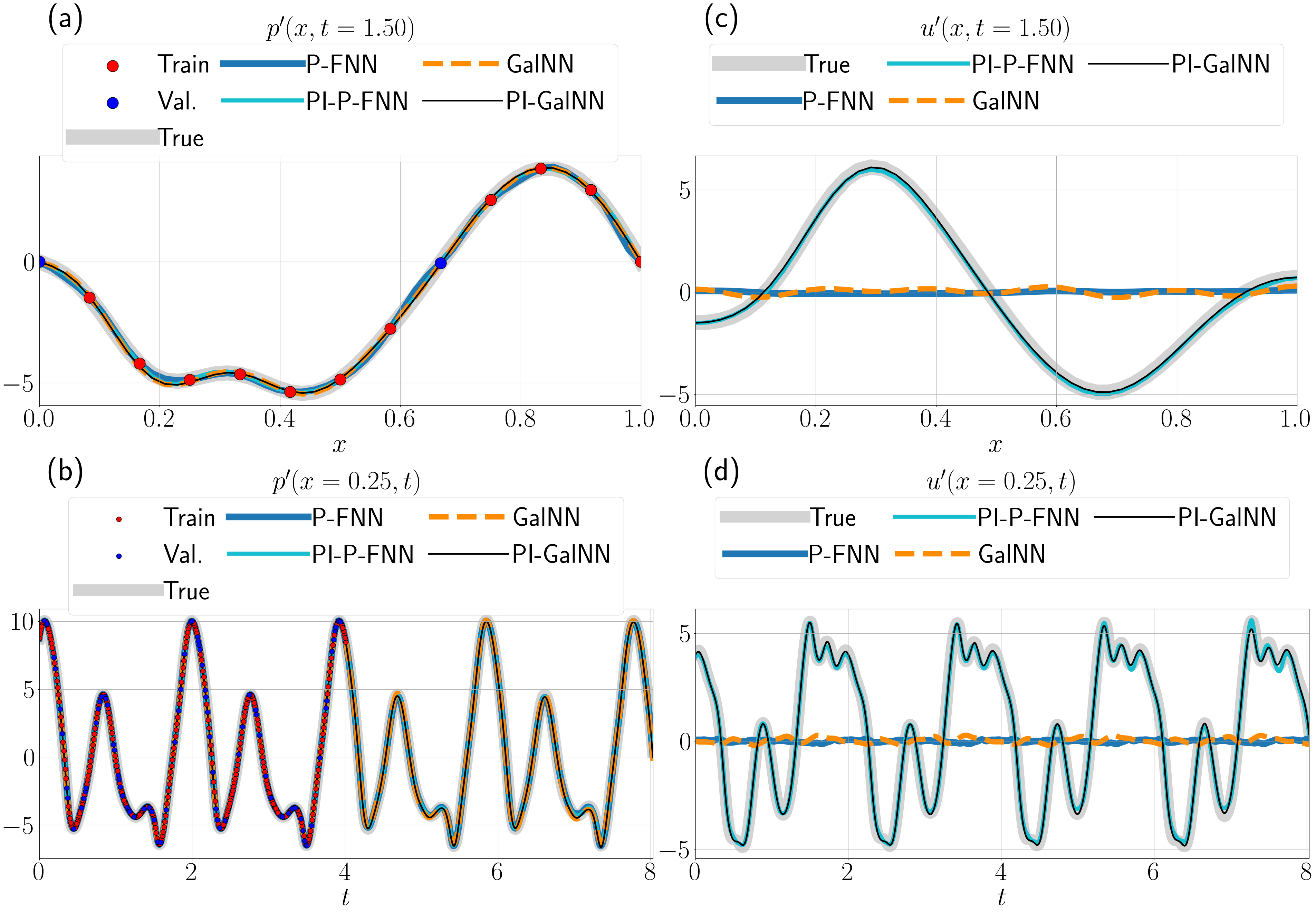}
    \caption{Reconstruction of acoustic velocity from pressure measurements only. Predictions on acoustic  (a, b) pressure and  (c, d) velocity from only 13 pressure observations with different physics-constrained networks.  
    Top row: spatial shapes at $t=1.5$.
    Bottom row: timeseries at $x=0.25$.
    \rev{The acoustic velocity is accurately reconstructed from pressure measurements with the physics-informed periodic feedforward and Galerkin neural networks.}}
    \label{fig:rijke_partial_obs}
\end{figure}
Full state measurements are difficult to perform in thermoacoustics, i.e., only pressure measurements are taken with microphones. To emulate such a scenario, we perform the training using only pressure data, discarding the velocity data. Therefore, in the formulation of the data-driven loss\eqref{eq:loss_data}, the measurement matrix becomes $\bm{M} = \begin{bmatrix}
    1 & 0 \\ 0 & 0 \\
\end{bmatrix}$. Our aim is to learn a model that can also output velocity information without ever being exposed to measurements from it during training. Figure \ref{fig:rijke_partial_obs} shows the predictions of the P-FNN and \rev{the GalNN} trained with and without physics-information. When the loss is computed solely from the data, only the pressure can be learned. On the other hand, velocity predictions are made possible by enforcing the physics via the physics-informed loss term. For the purely data-driven networks, the output weights related to velocity are not updated, hence the velocity predictions are made with the initial weight values which are small. \rev{When both momentum and energy equations are imposed, both physics-informed networks can accurately reconstruct the velocity.}

\section{Results on higher-fidelity data}\label{sec:results_higher}
\subsection{Higher-fidelity data}\label{sec:higher_model}
In order to assess how well the GalNN approach generalizes to higher-fidelity data, which is not generated by the Galerkin method, we employ a higher-fidelity model that includes the effects of the mean flow and has a kinematic flame model for the flame \citep{Dowling1999} (Figure \ref{fig:rijke}). 
The solution of the PDEs \eqref{eq:dim_pde_prime} can be obtained by a travelling wave approach in conjunction with the jump conditions at the flame. This model can also account for nonideal open boundary conditions via acoustic reflection coefficients $R_{in}$ and $R_{out}$ at the inlet and outlet. We simulate a system with ideal boundary conditions, i.e., $R_{in} = R_{out} = -1$, that results in limit-cycle oscillations using the mean flow parameters for the inlet velocity $\tilde{\bar{u}}_1 = 4 \: m/s$; for the inlet temperature $\tilde{\bar{T}}_1 = 293 \: K$; for the inlet and outlet pressure $\tilde{\bar{P}}_1 = \tilde{\bar{P}}_2 = 101300 \: Pa$; and for the heat release rate $\tilde{\bar{Q}} = 2000 \: W$. This is a low Mach number configuration with $\bar{M}_1 = 0.0117$ and $\bar{M}_2 = 0.0107$. \rev{The cross-sectional area changes at the flame location with a ratio of $A_1 / A_2 = 0.75$.} 
In the preprocessing of the data, we non-dimensionalize the variables as, $x = \tilde{x}/\tilde{L}$; $t = \tilde{t}\tilde{\bar{c}}_{ref}/\tilde{L}$;  $u = \tilde{u}/\tilde{\bar{c}}_{ref}$; $\rho = \tilde{\rho}/\tilde{\bar{\rho}}_{ref}$; $p = \tilde{p}/(\tilde{\bar{\rho}}_{ref}\tilde{\bar{c}}_{ref}^2)$; $\dot{q} = \tilde{\dot{q}}(\gamma-1)/(\tilde{\bar{\rho}}_{ref}\tilde{\bar{c}}_{ref}^3)$ where $\tilde{\bar{(\cdot)}}_{ref} = ({\tilde{L}_1\tilde{\bar{(\cdot)}}_1+\tilde{L}_2\tilde{\bar{(\cdot)}}_2})/{\tilde{L}}$ is chosen as the weighted average of the mean values of variable $(\cdot)$ across the flame.
The non-dimensional flame location is at $x_f = 0.61$, and the non-dimensional densities in the duct segments before and after the flame are $\bar{\rho}_1 = 1.145, \bar{\rho}_2 = 0.769$. These quantities will be used to determine the acoustic mode shapes for the Galerkin neural network \eqref{eq:rijke_galerkin_modes}. While creating the training and validation sets, we consider two types of validation; (i) validation of the interpolation capability, and (ii) validation of the extrapolation capability, i.e., when the data is sampled from outside the time range of training. A good validation of the interpolation loss prevents overfitting. Validation of extrapolation loss indicates how well the model generalizes beyond the time range of training. After discarding a transient, we split the first 4 non-dimensional time units long time series (approximately two periods) of the limit-cycle into training and validation of the interpolation sets, assigning a 80\% to 20\% ratio between them. The validation data for the extrapolation is then taken as the next 4 non-dimensional time units. 
The physics-informed loss of the higher-fidelity system is calculated by evaluating the left-hand side of the PDEs \eqref{eq:dim_pde_prime} after being non-dimensionalized as described above. The momentum residual, $\mathcal{F}_M$, is defined as the left-hand side of \eqref{eq:dim_momentum_prime}. In contrast to the twin experiments with Rijke tube, for the higher-fidelity data, we assume that we do not have access to the \rev{heat release model}. Hence, the energy residual, $\mathcal{F}_E$, is defined as the left-hand side of \eqref{eq:dim_energy_prime} excluding the heat release term, $\dot{q}$. This is acceptable because the heat release term only acts at the flame location due to the compact assumption and the Dirac delta.

\subsection{Comparison of networks performance for reconstruction} \label{sec:comparison_nns}
\begin{figure}[t!]
    \centering
    \includegraphics[width = \linewidth]{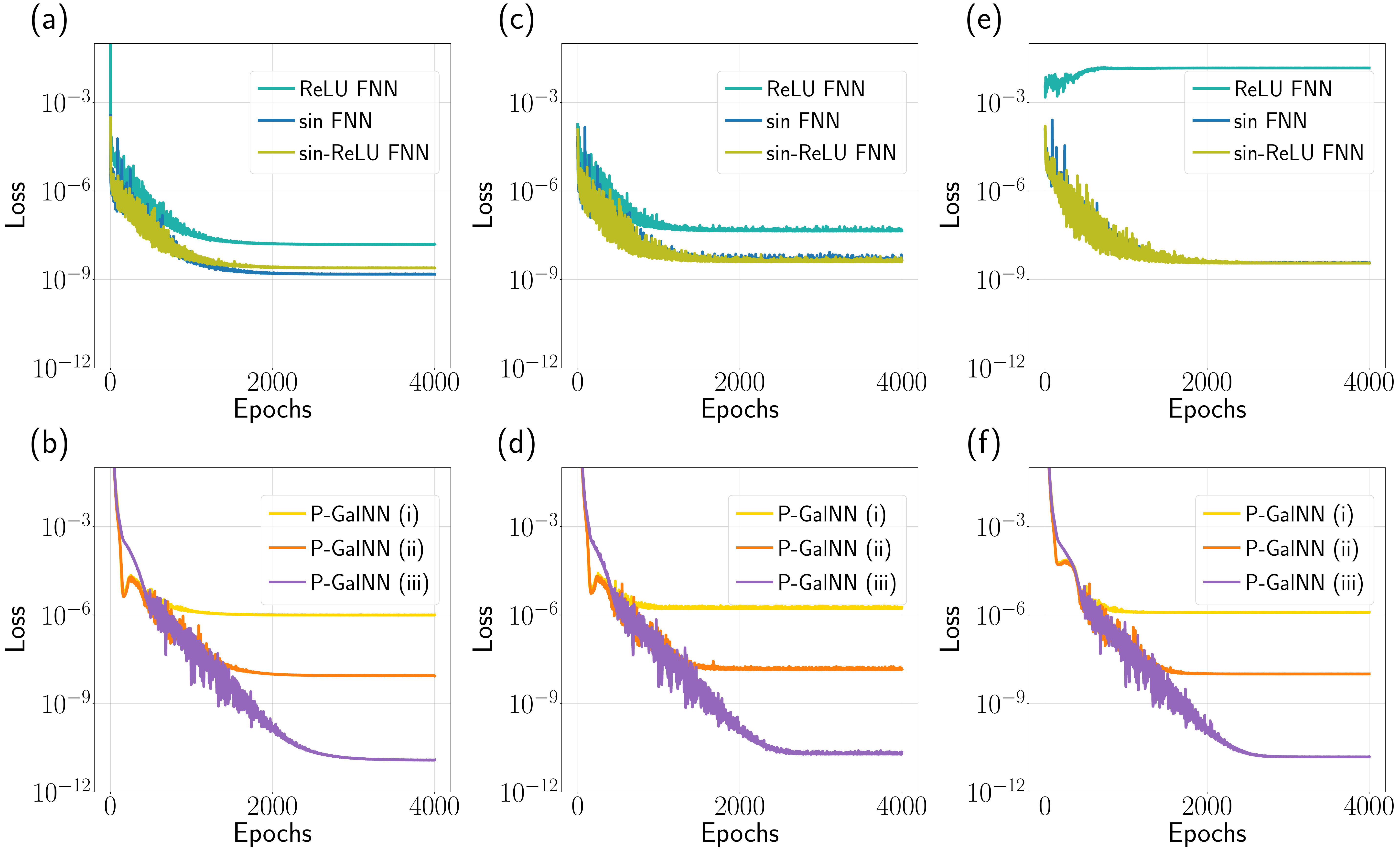}
    \caption{Data from higher-fidelity model for a periodic acoustic solution. Loss-function histories. (a-b) Training, (c-d) validation of interpolation, and (e-f) validation of extrapolation. Feedforward neural networks (FNNs) with different activations; ReLU, sine, sine-ReLU (sine in the first layer, ReLU in the rest). Period Galerkin neural networks (P-GalNNs) with different choices of spatial bases; (i) natural acoustic modes \eqref{eq:rijke_galerkin_modes} for $\bar{\rho}_1 = \bar{\rho}_2 = 1$, (ii) natural acoustic modes for $\bar{\rho}_1 = 1.145, \bar{\rho}_2 = 0.769$; and (iii) case (ii) with the addition of step functions in the velocity modes to handle the jump discontinuity \eqref{eq:discontinuity_modes}.}
    \label{fig:kinematic_loss_history}
\end{figure}
\begin{table*}[t!]
    \centering
    \caption{Summary of neural network model properties trained on higher-fidelity data. Feedforward neural networks (FNNs) with different activations; ReLU, sine, sine-ReLU (sine in the first layer, ReLU in the rest). Periodic Galerkin neural networks (P-GalNNs) with different choices of spatial bases;  (i) natural acoustic modes \eqref{eq:rijke_galerkin_modes} for $\bar{\rho}_1 = \bar{\rho}_2 = 1$, (ii) natural acoustic modes for $\bar{\rho}_1 = 1.145, \bar{\rho}_2 = 0.769$; and (iii) case (ii) with the addition of step functions in the velocity modes to handle the jump discontinuity \eqref{eq:discontinuity_modes}.}
    \begin{tabular}{c|c|c|c|c|c|c}
         & ReLU FNN & sin FNN & sin-ReLU FNN &  P-GalNN (i) & P-GalNN (ii) & P-GalNN (iii) \\
        \# hidden layers & 6 & 3 & 4 & 1 & 1 & 1 \\
        \# Galerkin modes & - & - & - & 20 & 20 & 20 \\
        Activations & ReLU & sine & sine-ReLU & harmonics & harmonics & harmonics \\
        \# neurons & 64 & 64 & 32 & 20 & 20 & 20 \\
        Learning rate & 0.01 & 0.01 & 0.01 & 0.0004 & 0.0004 & 0.0004 \\
        Optimizer & \multicolumn{6}{c}{Adam} \\
        Batch size & \multicolumn{6}{c}{32}
    \end{tabular}
    \label{tab:kinematic_nns}
\end{table*}
\begin{figure*}[t!]
    \centering
    \includegraphics[width = \linewidth]{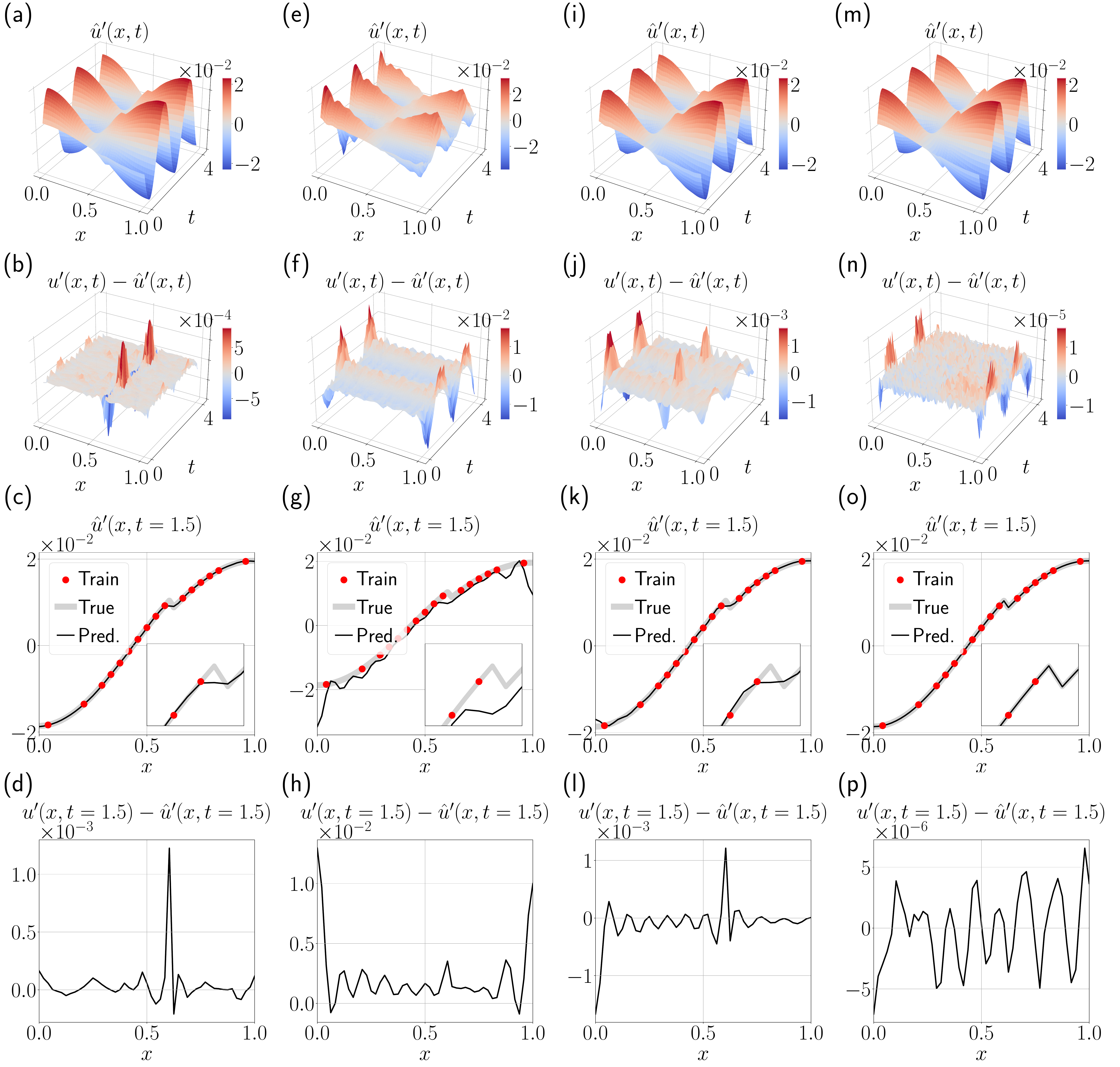}
    \caption{Comparison of acoustic neural networks in terms of reconstruction performance from higher-fidelity data. 
    From left to right, (a-d) sine-ReLU feedforward neural network (FNN), i.e., sine in the first layer, ReLU in the rest); 
    (e-h) periodic Galerkin neural network (P-GalNN) (i) with natural acoustic modes \eqref{eq:rijke_galerkin_modes} for $\bar{\rho}_1 =\bar{\rho}_2 = 1$; 
    (i-l) P-GalNN (ii) with natural acoustic modes for $\bar{\rho}_1 = 1.145, \bar{\rho}_2 = 0.769$; (m-p) P-GalNN (ii) with the addition of step functions in the velocity modes to handle the jump discontinuity \eqref{eq:discontinuity_modes}. From top to bottom, velocity prediction in time and space, error between ground truth and prediction in time and space, velocity prediction and error at a fixed time instance, $t = 1.5$.}
    \label{fig:fnn_gnn_grid_plot}
\end{figure*}

We analyse the performance of the different networks that we have built in Section \ref{sec:acoustic_nn} on the reconstruction of the higher-fidelity data. We compare three configurations of the feedforward neural network with; (i) ReLU activations in all hidden layers, (ii) sine activation in all hidden layers, (iii) sine activation in the first layer, ReLU activation in the rest of the hidden layers, and three configurations of the periodic Galerkin neural network with one hidden trainable frequency layer; (i) with $\omega_j = j\pi$ as the acoustic angular frequencies as before, i.e., when $\bar{\rho}_1 = \bar{\rho}_2 = 1$, (ii) with $\omega_j$ determined from the dispersion relationship using the real mean flow density values of the system, i.e., when $\bar{\rho}_1 = 1.145, \bar{\rho}_2 = 0.769$, (iii) the Galerkin modes chosen as (ii) with the addition of step functions in the velocity modes to handle the jump discontinuity \eqref{eq:discontinuity_modes}. 

For each configuration of activations, the hyperparameters of the architecture and training (number of layers, number of neurons, and learning rate) have been tuned with a grid search, in which we chose the set of hyperparameters that resulted in the smallest validation losses. These properties are provided in Table \ref{tab:kinematic_nns} for each network.
The training and validation loss histories of all models are shown in Figure \ref{fig:kinematic_loss_history}. The ReLU network has higher training and interpolation losses, cannot fit the data as well as the sine and sine-ReLU networks, and cannot extrapolate. For all the other networks (sine FNN, sine-ReLU FNN, P-GalNN (i), P-GalNN (ii), P-GalNN (iii)), the validation losses are close to the training loss. The P-GalNN (iii) scores the lowest losses. The periodic Galerkin networks can have similar or better prediction performance as the other deeper feedforward neural networks even with one hidden layer following a physical choice of spatial basis. Figure \ref{fig:fnn_gnn_grid_plot} visualizes the velocity predictions and their errors between the predictions and the true data obtained with the networks. 
The P-GalNN (i) 
exhibits an error pattern that resembles an offset from the true solution, which indicates a biased model. 
The prediction is also oscillatory and the error is especially high at the boundaries. The more physically motivated choice of P-GalNN (ii)
eliminates these errors, however the discontinuity jump still can not be captured, which leads to high frequency oscillations in the prediction around the flame location. A similar phenomenon is also observed for the sine-ReLU network, where the error is the highest at the flame location. This problem is overcome by 
P-GalNN (iii), which results in the smallest training and validation errors overall. This network finds the non-dimensional dominant angular frequency as $\theta^* = 3.3164$ and thus, the period as $T^* = 1.8946$. From the power spectral density of the true pressure signal, we read $\theta^* = 3.3166$ and $T^* = 1.8945$. The periodic Galerkin network successfully learns the correct period.

\subsection{Robustness to sparsity and noise}
\begin{figure*}[t!]
    \centering
    \includegraphics[width = \linewidth]{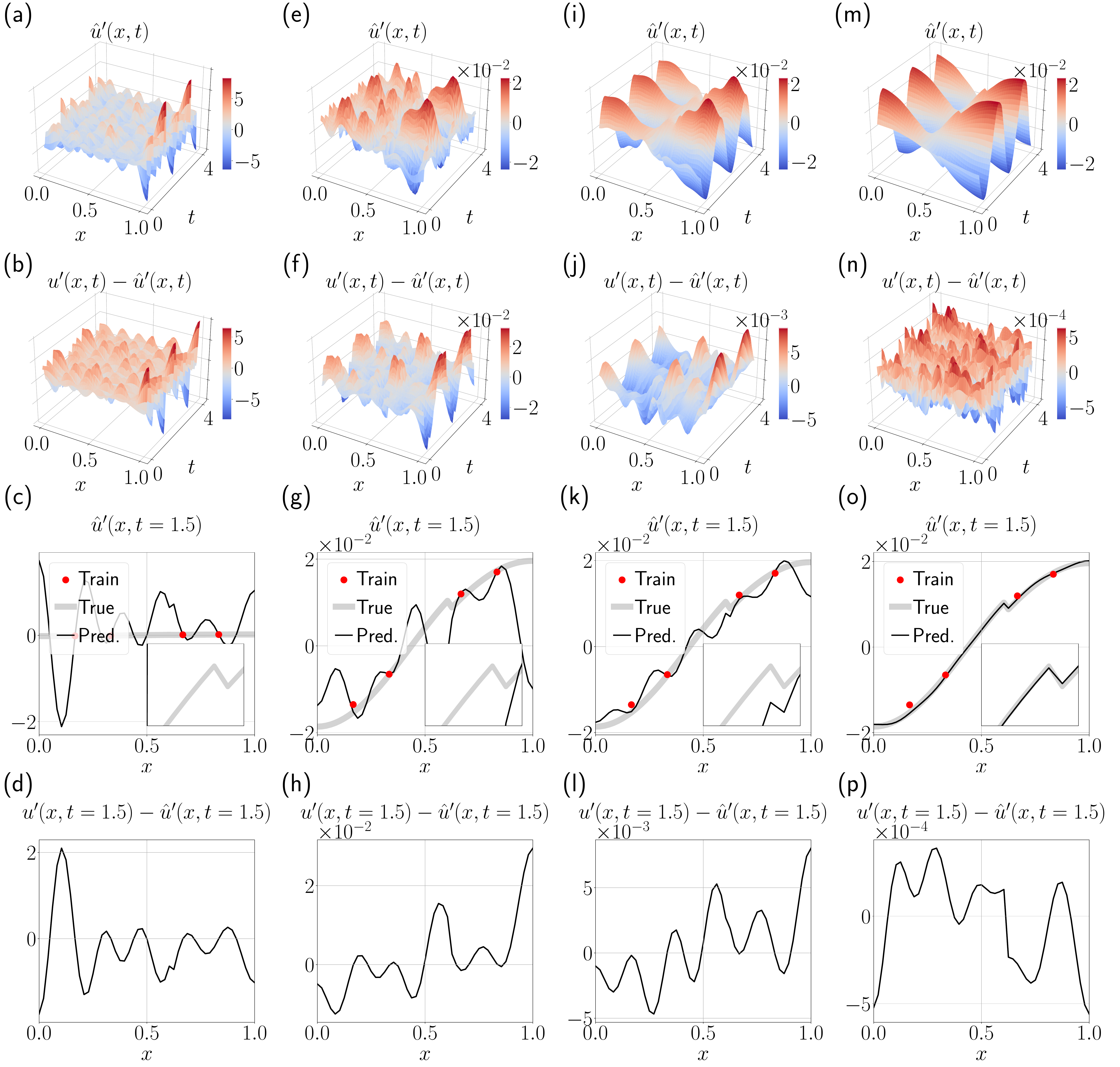}
    \caption{In reference to Fig.~\ref{fig:fnn_gnn_grid_plot}, effect of the regularization hyperparameter on periodic Galerkin neural networks (P-GalNNs) in case of sparse and noisy measurements. From left to right, (a-d) no regularization, (e-h) $\ell_2$-norm regularization, (i-l) $\ell_1$-norm regularization, and (m-p) physics-based regularization. From top to bottom: acoustic velocity, error between ground truth and prediction, velocity prediction at a fixed time instance, and error at a fixed time instance, $t = 1.5$.}
    \label{fig:gnn_reg_grid_plot}
\end{figure*}

\begin{figure}[t!]
    \centering
    \includegraphics[width = 0.8\linewidth]{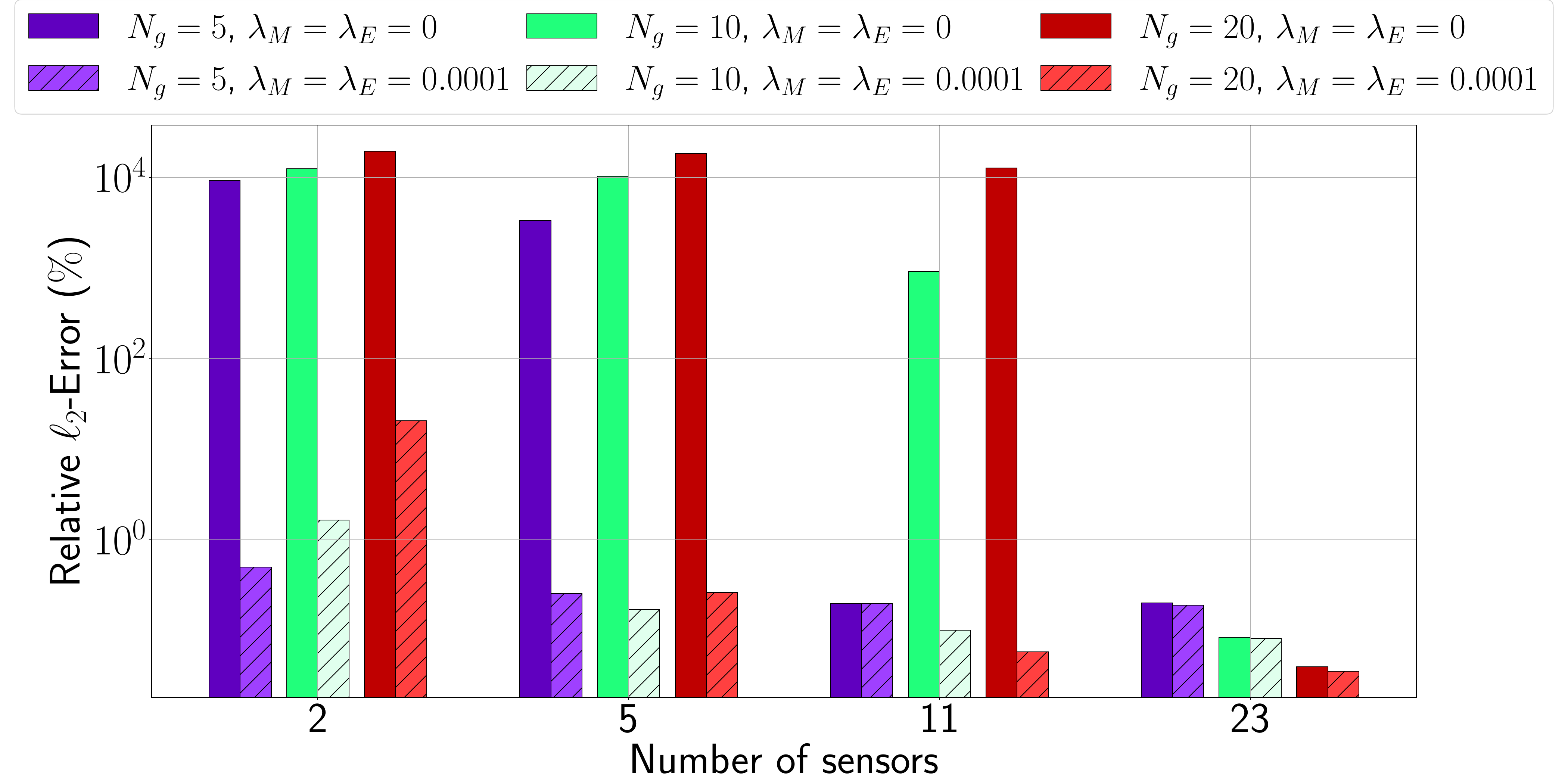}
    \caption{Effect of number of pressure sensors in data collection. Effect of number of modes in Galerkin neural networks, with and without physics-based loss. (The regularization hyperparameter of the data-driven loss is $\lambda_{D} = 1$.)}
    \label{fig:robustness_sensor}
\end{figure}

\begin{figure}[t!]
    \centering
    \includegraphics[width = 0.7\linewidth]{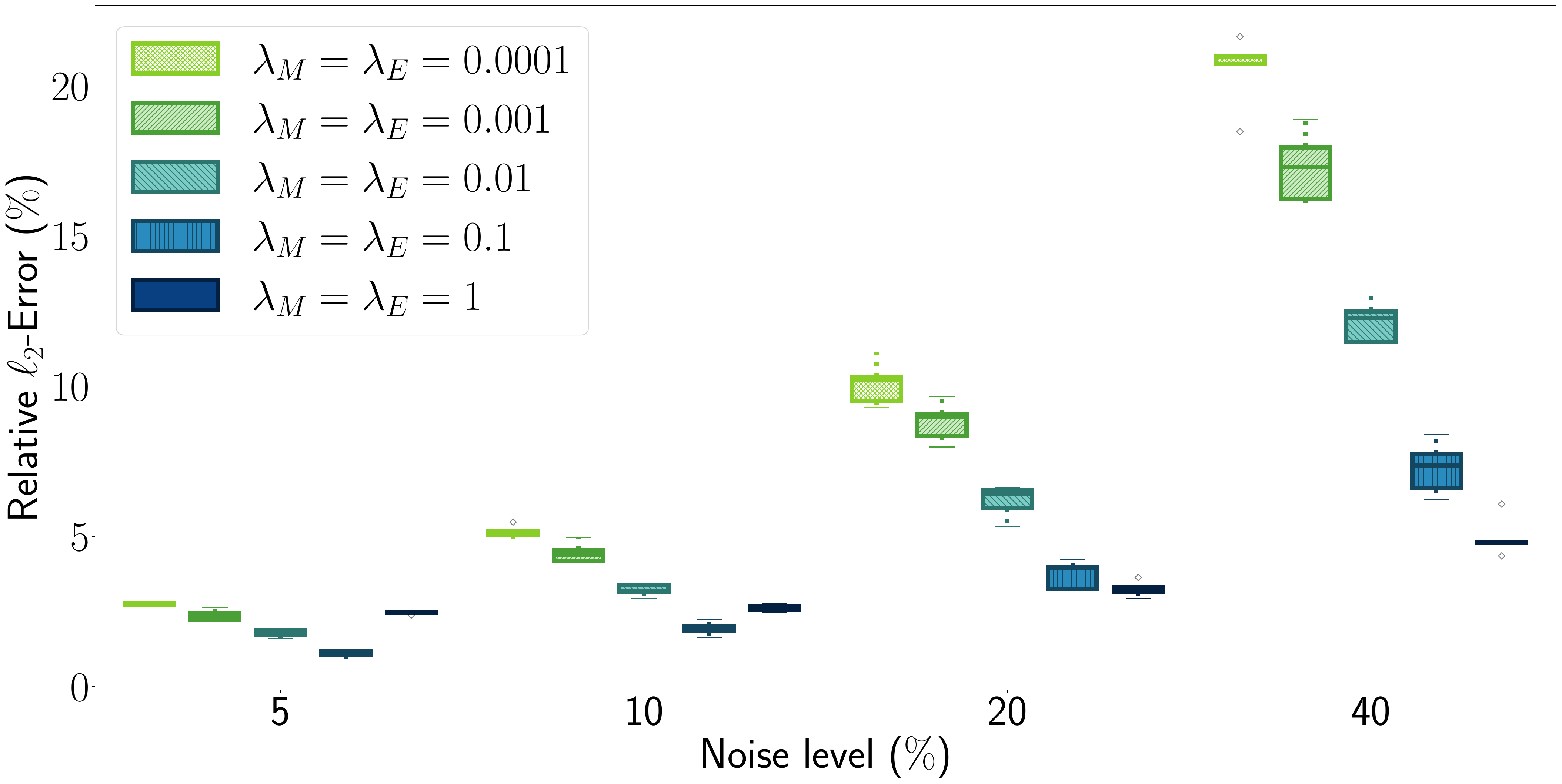}
    \caption{Robustness to noise. Galerkin neural networks with different regularization hyperparameters for physics-based losses with data corrupted with noise. (The regularization hyperparameter of the data-driven loss is $\lambda_{D} = 1$.)}
    \label{fig:robustness_noise}
\end{figure}

\begin{figure}[t!]
    \centering
    \includegraphics[width = \linewidth]{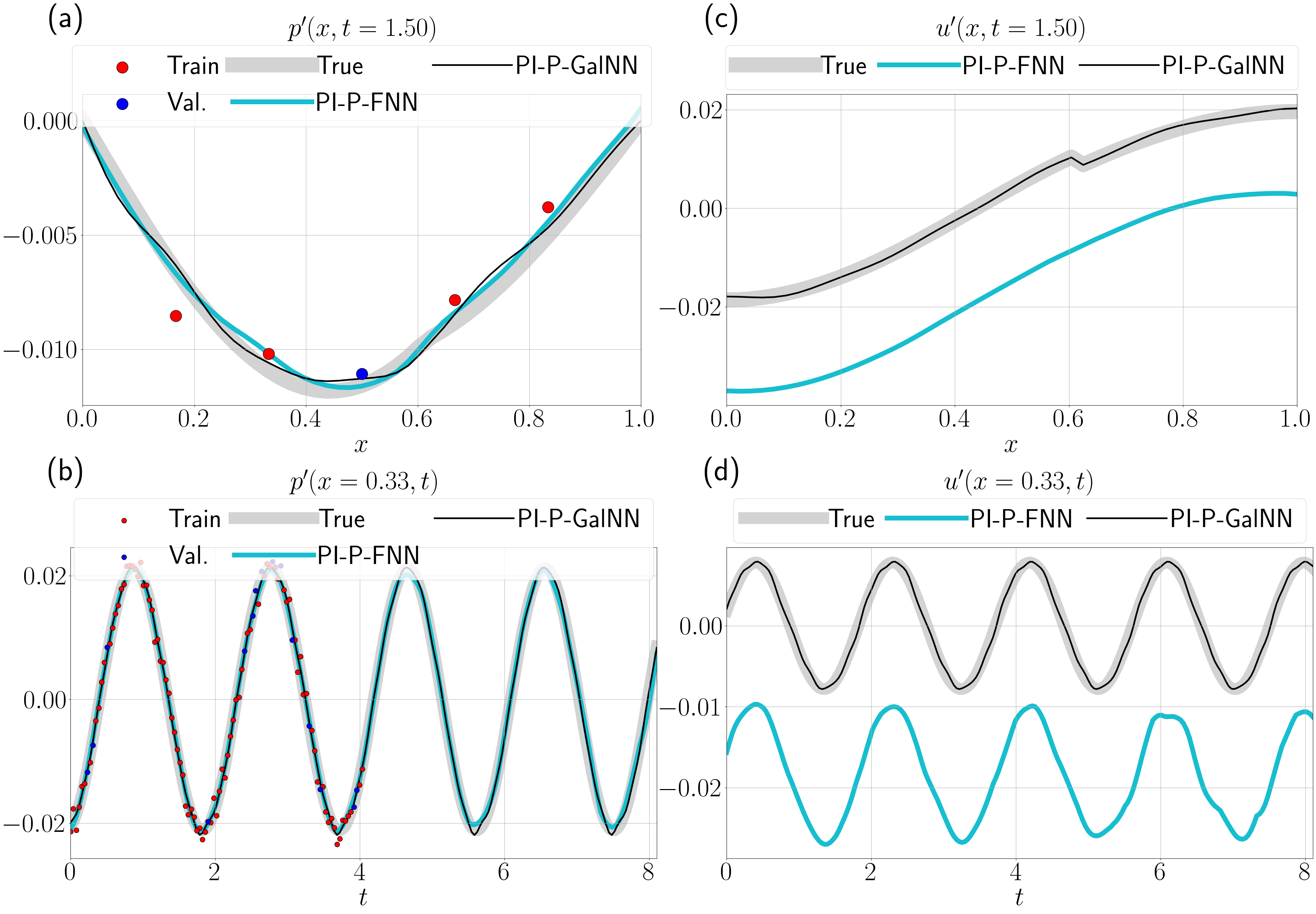}
    \caption{
    Reconstruction of acoustic velocity from noisy pressure measurements only. Predictions on acoustic  (a, b) pressure and (c, d) velocity from only 5 pressure observations with physics-based loss using the periodically activated feedforward neural network (P-FNN) and periodic Galerkin neural network (P-GalNN). Top row: pressure and velocity shapes along the tube at a fixed time instance, $t = 1.5$. Bottom row: pressure and velocity timeseries at a fixed point in the tube, $x = 0.33$.}
    \label{fig:kinematic_partial_obs}
\end{figure}
In this section, we demonstrate the robustness of the acoustic neural networks to sparsity and noise in the training data. We select P-GalNN (iii) from the previous section because of its more generalizable performance in comparison to the other architectures and investigate the regularization methods that can be used in order to prevent overfitting in the case of corrupted or scarce data. We compare a physics-based regularization, i.e., physics-informed loss, with $\ell_1$- and $\ell_2$-norm regularizations, which act on the network weights as $||\bm{W}||_1$ and $||\bm{W}||_2$ respectively in the loss term. 
We add zero mean Gaussian noise on the non-dimensional training data with a standard deviation equal to 10\% of the standard deviation of the true solution function over the whole domain \rev{(20 dB signal-to-noise ratio)}. Using the described regularization methods, we train networks with 10 Galerkin modes on this new dataset this time given only on a coarse grid. The comparison of the obtained predictions when using (i) no regularization, (ii) $\ell_2$-norm, (iii) $\ell_1$-norm, and (iv) physics-based regularization is shown in Figure \ref{fig:gnn_reg_grid_plot}. 
We train multiple networks with varying values of regularization coefficients and show only the results for those with the smallest validation losses. Without regularization, the model can fit the training data perfectly, however the interpolation points show spikes, which is a clear demonstration of overfitting. This case of overfitting emerges when there is a discrepancy between the number of sensors and the number of Galerkin modes dictated by Nyquist-Shannon sampling theorem. The highest number of Galerkin modes is given as 
\begin{equation}
  N_g^* = \underset{j}{\mathrm{arg \, max}}\; \{j \; | \; \omega_j\sqrt{\bar{\rho}_1} \leq k_{Nyquist}, \; \omega_j\sqrt{\bar{\rho}_2} \leq k_{Nyquist}\}  
\end{equation}
where $k_{Nyquist} = \frac{\pi}{\Delta x}$ with $\Delta x$ being the spacing between sensor locations. 
The physics information significantly eliminates overfitting and outperforms $\ell_1$- and $\ell_2$-norm regularizations. The $\ell_2$-norm regularization does not work as well because the weights are not selectively regularized, i.e., dominant modes/frequencies will be regularized in the same way as non-dominant ones. The $\ell_1$-norm regularization performs better since modes corresponding to higher wavenumbers are discarded as $\ell_1$-norm promotes sparsity and in the physical basis, the energy is contained in the low frequency modes. 

We perform a quantitative study to investigate the robustness of the prediction performance to sparsity and noise. First, we vary the number of sensors along the tube. Figure \ref{fig:robustness_sensor} shows the relative $\ell_2$-error achieved by Galerkin networks with different number of modes when trained on noise-free data given on spatial grids of 2, 5, 11, and 23 points. 
The relative $\ell_2$-error is calculated from the prediction on the fine grid of 49 points over the training range by dividing the $\ell_2$-norm of the error by the $\ell_2$-norm of the ground truth. 
The networks are trained with and without physics-information in the loss function. We fix the weighting of the data-driven loss to $\lambda_{D} = 1$, and the physics loss to $\lambda_{M} = \lambda_{E} = 0.0001$ for physics-informed and to 0 otherwise. As discussed above, high number of Galerkin modes leads to overfitting for coarse grids, when the Nyquist wavenumber is not sufficient to resolve the high wavenumbers, e.g., 5 sensors and 10 Galerkin modes. Furthermore, we observe large prediction error even when the Nyquist condition is satisfied, e.g. 5 sensors and 5 Galerkin modes. We find that this type of error is mostly concentrated at the boundaries of the velocity, as we have not provided any boundary data and therefore, the model overfits to the rest of the data, not generalizing well over the boundaries. For pressure, this is not an issue, because the boundary conditions are Dirichlet and already encapsulated within the provided Galerkin basis. On the other hand, low number of Galerkin modes may not be enough to approximate the pressure and velocity shapes in fine detail, e.g., 11 sensors and 5 Galerkin modes. The addition of the physics-information shows a marked difference and helps overcoming these limitations. \rev{In Appendix \ref{sec:partial_sensor}, we show the reconstruction errors associated with physics-informed training with varying numbers of pressure sensors.}

Next, we choose the network with 10 modes and add increasing levels of noise to the data from 5 sensor measurements. We take 5 different realizations of noise and show the prediction relative $\ell_2$ error as a box plot in Figure \ref{fig:robustness_noise} for noise levels of 5, 10, 20, and 40 \% \rev{(26, 20, 14, 8 dB signal-to-noise ratios)}. Noting that even in the noise-free case, the non-physics-informed network performs poorly, we focus on physics-informed training with varying weightings of the physics loss. As the noise level increases, the optimum weighting of the physics-information increases as well, since the data becomes less reliable. We obtain good predictions of pressure and velocity over the entire domain, even though the data is sparse and noisy. The network is also robust to different realizations of noise.

We conclude by demonstrating how the physics-informed training performs in the absence of velocity data. 
The prediction results of a physics-informed periodic Galerkin neural network, PI-P-GalNN, are compared with a sine-ReLU physics-informed feedforward neural network, PI-P-FNN, in Figure \ref{fig:kinematic_partial_obs}. For the PI-P-GalNN, we report the following relative $\ell_2$-errors: 
2.63 $\%$ for pressure and 2.75 $\%$ for velocity in the training range ($t = 0-4$ time units), and 2.84 $\%$ for pressure and 2.80 $\%$ for velocity in the extrapolation range ($t = 4-8$ time units). 
In comparison, for the sine-ReLU feedforward neural network, we report the following relative $\ell_2$-errors: 4.81 $\%$ for pressure and 164.22 $\%$ for velocity in the training range ($t = 0-4$ time units), and 9.77 $\%$ for pressure and 160.57 $\%$ for velocity in the extrapolation range ($t = 4-8$ time units). With the PI-P-GalNN, we can reconstruct the velocity as well as pressure. The PI-P-FNN fits to the noisy data more compared to the PI-P-GalNN, and can only recover the velocity with an offset, completely discarding the velocity jump. \rev{A constant offset of the velocity, $\hat{u}'(x,t) = u'(x,t) + c$, satisfies the governing equations \eqref{eq:dim_pde_prime}. Because the partial derivatives of the constant offset will be zero, the residual of the governing equations will not be affected. If the velocity fluctuations are known to have a zero mean,  one strategy could be to remove the mean of the predictions in the postprocessing, and a second strategy could be to remove the biases from the output layer of the neural network. However, even when the offset is removed, the PI-P-FNN can not capture the jump of the velocity at the flame location.}

\subsection{Nonideal boundary conditions}
\begin{figure}[t!]
    \centering
    \includegraphics[width=\linewidth]{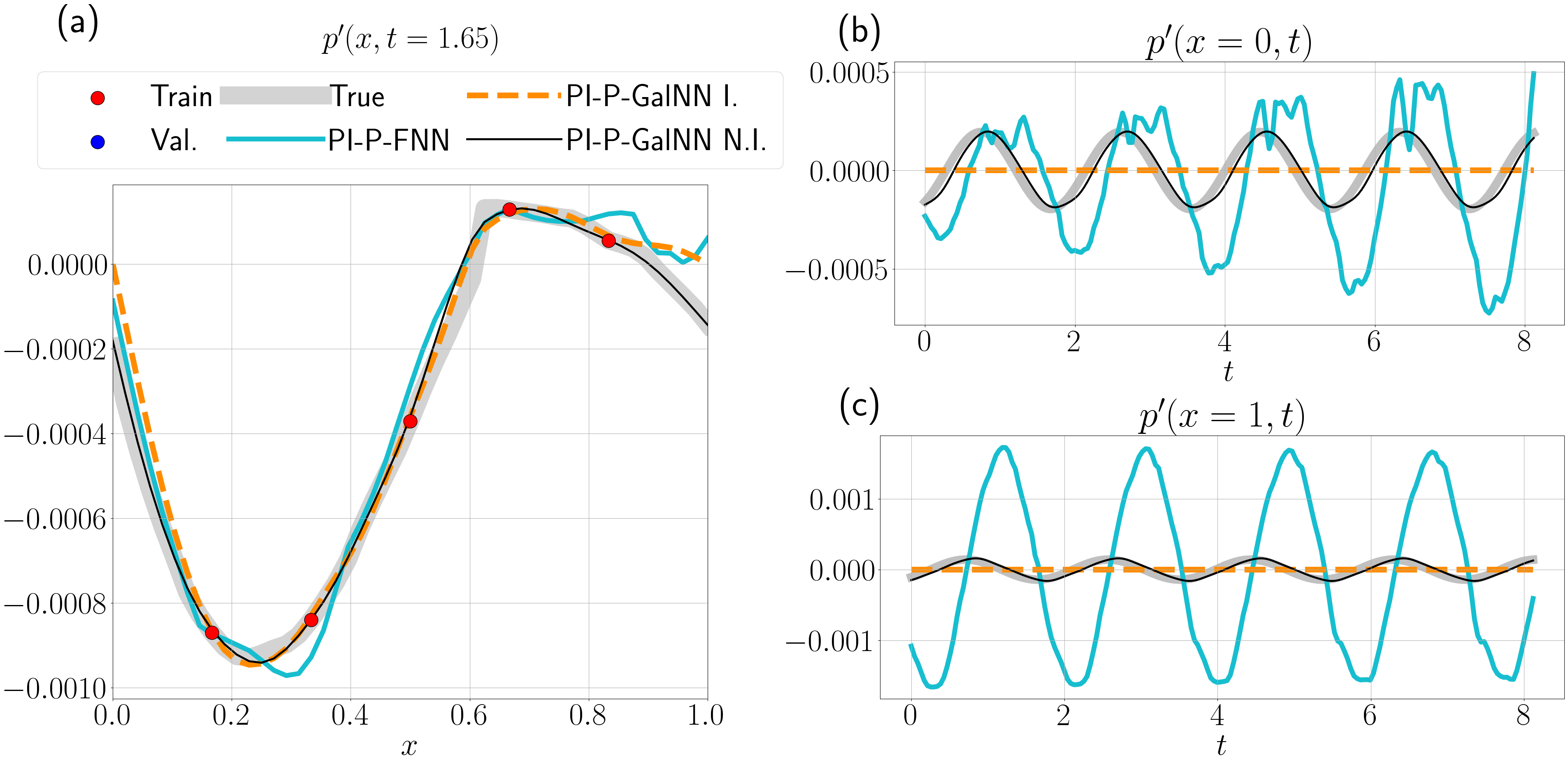}
    \caption{\rev{Prediction in the case of nonideal ($R_{in}=R_{out}=-0.985$) boundary conditions for a training case of only 5 noise-free pressure sensors. Comparison of a standard physics-informed periodic Galerkin neural (PI-P-GalNN I.), physics-informed periodic feedforward network (PI-P-FNN) with a PI-P-GalNN for nonideal boundary conditions (PI-P-GalNN N.I.). Shown in (a) pressure shape at a fixed time instance, $t=1.65$, (b,c) pressure fluctuations at $x=0$ and $x=1$, respectively. The linear term can capture the non-zero pressure fluctuations at the boundaries, when they are close to ideal.}}
    \label{fig:reflection}
\end{figure}

\begin{figure}[t!]
    \centering
    \includegraphics[width=\linewidth]{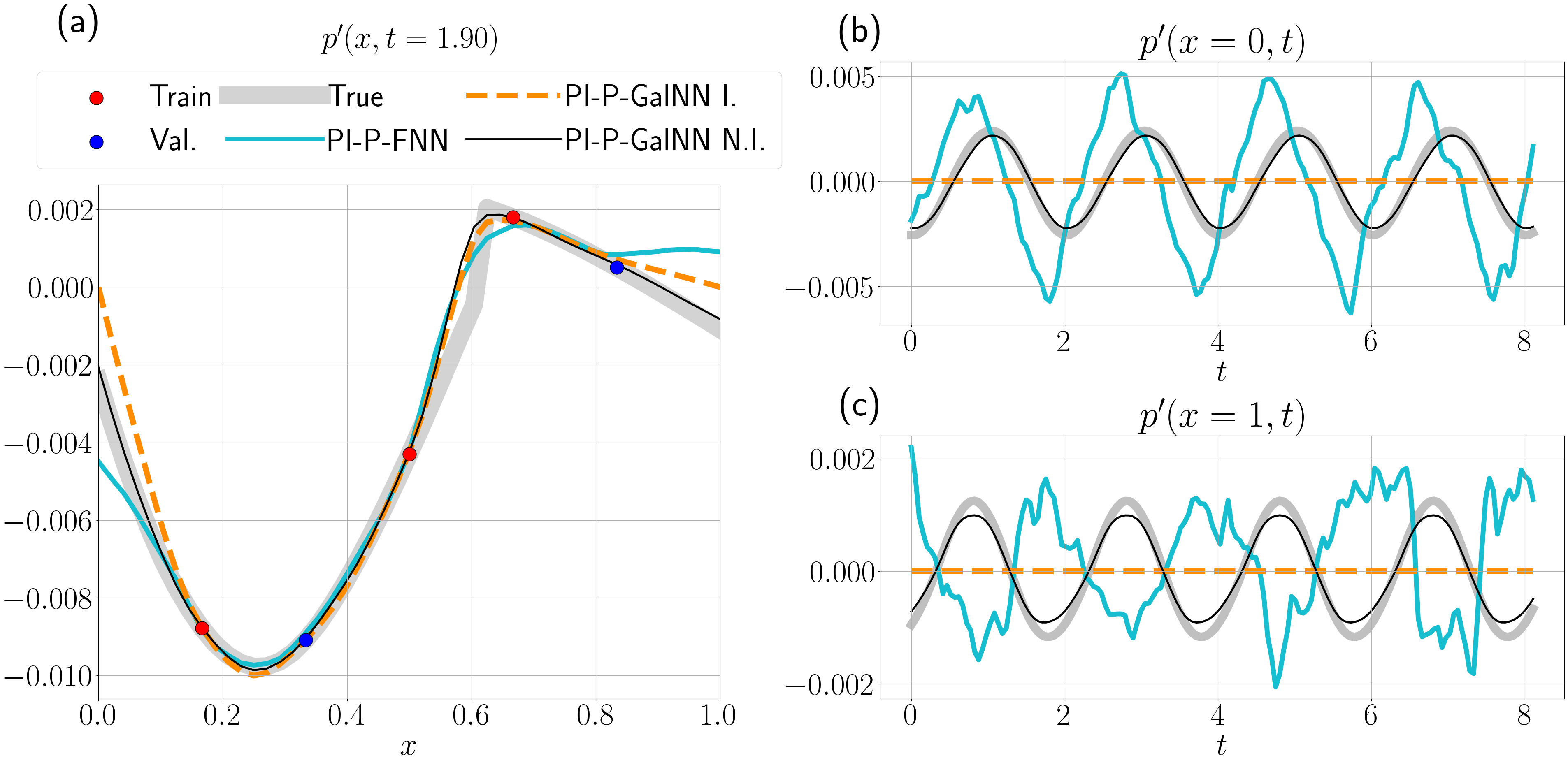}
    \caption{\rev{Prediction in the case of nonideal ($R_{in}=R_{out}=-0.85$) boundary conditions for a training case of only 5 noise-free pressure sensors. Comparison of a standard physics-informed periodic Galerkin neural (PI-P-GalNN I.), physics-informed periodic feedforward network (PI-P-FNN) with a PI-P-GalNN for nonideal boundary conditions (PI-P-GalNN N.I.). Shown in (a) pressure shape at a fixed time instance, $t=1.90$, (b,c) pressure fluctuations at $x=0$ and $x=1$, respectively. The linear term can capture the non-zero pressure fluctuations at the boundaries, when they are less close to ideal.}}
    \label{fig:reflection2}
\end{figure}

\begin{figure}[t!]
    \centering
    \includegraphics[width=\linewidth]{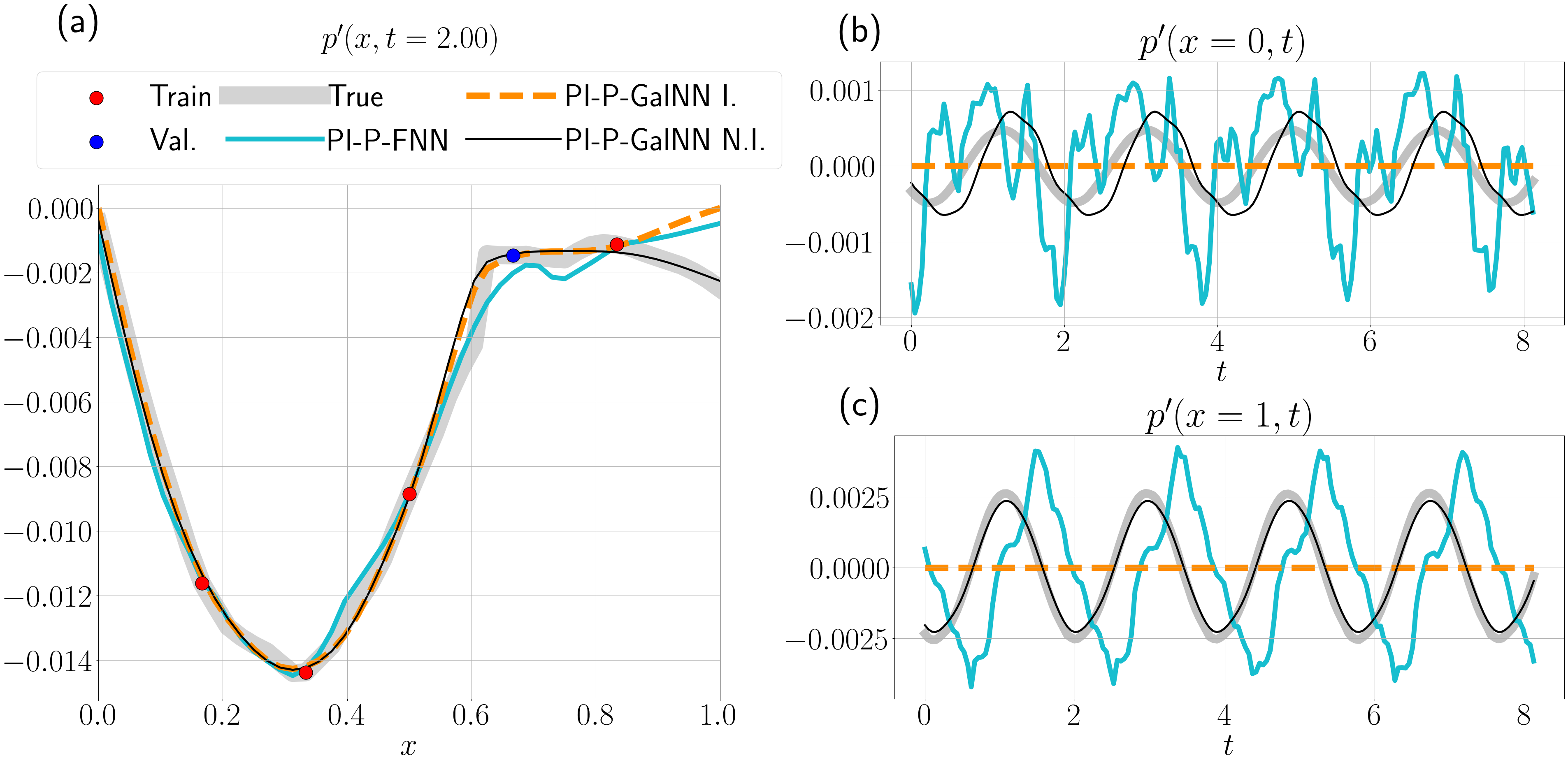}
    \caption{\rev{Prediction in the case of nonideal boundary conditions ($R_{in}=-0.985, R_{out}=-0.85$) for a training case of only 5 noise-free pressure sensors. Comparison of a standard physics-informed periodic Galerkin neural (PI-P-GalNN I.), physics-informed periodic feedforward network (PI-P-FNN) with a PI-P-GalNN for nonideal boundary conditions (PI-P-GalNN N.I.). Shown in (a) pressure shape at a fixed time instance, $t=2.00$, (b,c) pressure fluctuations at $x=0$ and $x=1$, respectively. The linear term can capture the non-zero pressure fluctuations at the boundaries, even when the reflection coefficients are different.}}
    \label{fig:reflection3}
\end{figure}
So far, we have restricted ourselves to ideal boundary conditions. However, in real experiments, we deal with nonideal conditions, in which the acoustic waves are not fully reflected at the tube ends ~\citep[e.g.,][]{Levine1948}. As a result, in contrast to the ideal case and the acoustic modes we have obtained under that assumption, the pressure fluctuations at the tube ends are not zero, i.e., $p'(x=0,t) \neq 0, \; p'(x=1,t) \neq 0$. In order to tackle nonideal boundary conditions, we add a linear term in the pressure modes \eqref{eq:pi_mode},
\begin{subequations}
\begin{align}
     \Pi_{N_g+1}^{(1)}(x) &= \Pi_{N_g+1}^{(2)}(x) = x, \\
    \Pi_{N_g+2}^{(1)}(x) &= \Pi_{N_g+2}^{(2)}(x) = 1, 
\end{align}
\end{subequations}
such that the summation \eqref{eq:rijke_galerkin_p} runs from 1 to $N_g+2$. The modes $\Pi_{N_g+1}$ and $\Pi_{N_g+2}$ are weighted by independent coefficients, $\mu_{N_g+1}$ and $\mu_{N_g+2}$. This way, the linear term provides the offset to the predicted pressure fluctuations at the tube ends, i.e., $\mu_{N_g+1}(t)x + \mu_{N_g+2}(t)$. Once trained, we expect to estimate $\mu_{N_g+2}(t) = p'(x=0,t)$ and $\mu_{N_g+1}(t) = p'(x=1,t)-p'(x=0,t)$.

\rev{We consider three cases and generate data from the \rev{higher-fidelity} model using the following parameters: 
(1) Close to ideal: $\tilde{\bar{u}}_1 = 8 \: m/s$, $\tilde{\bar{Q}} = 10000 \: W$, $R_{in}=R_{out}=-0.985$; 
(2) less close to ideal: $\tilde{\bar{u}}_1 = 16 \: m/s$, $\tilde{\bar{Q}} = 60000 \: W$, $R_{in}=R_{out}=-0.85$; and 
(3) different at both ends: $\tilde{\bar{u}}_1 = 16 \: m/s$, $\tilde{\bar{Q}} = 40000 \: W$, $R_{in}=-0.985, R_{out}=-0.85$}

\rev{The non-dimensional densities in the duct segments before and after the flame are for case (1) $\bar{\rho}_1 = 1.269, \bar{\rho}_2 = 0.571$, for case (2) $\bar{\rho}_1 = 1.434, \bar{\rho}_2 = 0.307$, and for case (3) $\bar{\rho}_1 = 1.376, \bar{\rho}_2 = 0.400$.}

\rev{Figures \ref{fig:reflection}, \ref{fig:reflection2}, and \ref{fig:reflection3} compare the predictions of a PI-P-GalNN (15 Galerkin modes) with a linear term against the standard PI-P-GalNN and the PI-P-FNN for the above cases, where the training data consists of only five pressure sensors. With the addition of the linear term, the PI-P-GalNN is capable of capturing the non-zero pressure fluctuations at the boundaries for near ideal conditions even from sparse measurements given at other locations. While PI-P-FNN is a flexible tool that can handle  various boundary conditions, it requires the boundary data \citep{silvagarzon2023ReconstructionAcousticFields}, wheres the PI-P-GalNN does not. (Training computational costs are provided in Appendix \ref{sec:costs}.)}

\section{Conclusions}
In this work, we model acoustic and thermoacoustic pressure and velocity oscillations from synthetic data. The synthetic data captures the rich nonlinear behaviour of thermoacoustic oscillations observed in propulsion and power generation. We develop acoustic neural networks to tackle the tasks of (i) extrapolation in space and time, and (ii) reconstruction of full acoustic state from partial observations by exploiting prior knowledge on the physics of acoustics. The prior knowledge is embedded in the network as both soft and hard constraints. 
First, as acoustic and thermoacoustic systems are dominated by sinusoidal eigenfunctions, we promote a physically-motivated choice of sinusoidal activation functions. Unlike standard feedforward neural network architectures that employ ReLU or tanh activations, periodically activated networks can extrapolate in time, and the trained weights hold frequency information. This means that spatiotemporal patterns can be learned more efficiently from less data because the network is more expressive as well as robust. 
Second, we inform the training with the acoustic conservation laws with penalty terms in the loss function. This term regularizes the predictions in the case of noisy and scarce data, and enables reconstruction of unobserved states. Typically, in thermoacoustics experiments, only pressure measurements are available.
Third, inspired by Galerkin decomposition, we design a neural network with temporal and spatial branches (Galerkin neural network), which spans a physical function space via the choice of acoustic eigenmodes as the spatial basis. In order to account for nonideal boundary conditions that occur when the acoustic waves are not fully reflected at the boundaries, we add a linear mode in the pressure modes, which captures the non-zero pressure fluctuations at the inlet and outlet for near ideal conditions without \rev{requiring data from the boundaries}. 
We consider two test cases: (i) twin experiments on synthetic data from a Rijke tube with a nonlinear, time-delayed \rev{heat release model} with Galerkin discretization of the PDEs, and (ii) higher-fidelity data generated by a thermoacoustic network model with a kinematic flame model that takes into account mean-flow effects. \rev{The first case shows that the standard feedforward neural networks fail at extrapolation, while periodically activated networks can extrapolate and reconstruct accurately from partial measurements.} On this test case, we also demonstrate the long term extrapolation capability of Galerkin neural networks on periodic and quasi-periodic solutions, while for chaotic solutions, we recover the dominant components in the frequency spectrum. In the second test case, we show the generalizability and robustness of the Galerkin neural networks to higher-fidelity data, which can be noisy and contain only pressure measurements.
%
\rev{Hard-constraining neural networks with prior knowledge reduces the  required network size, search space of hyperparameters as well as the amount of data required for training. }  This work opens up possibilities to learn the nonlinear dynamics of thermoacoustics using physics-aware data-driven methods. 
\rev{Future research directions include testing the method for geometries with cross-section changes, in which the predetermined spatial branch of the Galerkin neural network can be extended with more trainable parameters.}

The code is available at~\citep{github_repo}.
\begin{acknowledgments}
This research has received financial support from the ERC Starting Grant No. PhyCo 949388. \rev{We gratefully acknowledge an anonymous reviewer for their spot-on comments, which helped improved the paper.}
\end{acknowledgments}

\appendix
\section{Calculation of modal damping and heat release terms in the physics-informed loss}\label{sec:fnn_pi_appendix}
\subsection{Modal damping}
The modal damping is defined as a multiplication in the frequency domain and a convolution in the spatial domain with the pressure. Using an FNN, we predict the pressure at a given time in the spatial domain. In order to compute the contribution of the damping in the physics-informed loss, at each training step, we will transform the pressure predicted in the spatial domain to the frequency domain via Fast Fourier Transformation (FFT), compute the damping in the frequency domain, and then transform the result back to the spatial domain. When sampling from signals, the distinct Fourier frequencies are given as $\omega_k = k\frac{2\pi}{KT_s}, k = 0,1,...,K/2$, where $K$ is total number of samples, $T_s$ is the sampling time, and $KT_s$ gives the length of the sampled signal. During simulation, we considered $N_g$ Galerkin modes. So, in our case, we set $K \geq 2N_g$. Notice that the spatial domain is $[0,1]$, while the wavenumbers of the Galerkin modes have a resolution of $\pi$, which is only half-period in this domain. Consequently, the resolution of the Fourier frequencies must be set to $\frac{2\pi}{KT_s} = \pi$ and hence, $KT_s= 2$, which means that the FFT must be taken over a domain of $[0,2]$, such that the samples are at $x_k = k\frac{2}{K}, k = 0,1,..,K-1$ and we have one period of discrete samples. One could predict the pressure over the $[0,2]$ domain, but this could lead to errors as there is no training data in this region. Instead, we first predict over the original $[0,1]$ domain and then stack this prediction with its negative symmetric with respect $x = 1$, since we know that pressure is given as a sum of sines, which is an odd function. So, we have
\begin{equation}
    \hat{p}' = \left\{\hat{p}'(x_k, t) \: \middle\vert x_k = k\frac{2}{K}, \: k = 0,1,..,K-1,  \: K \geq 2N_g\right\},
\end{equation}
and its Fourier Transform $\mathcal{F}(\hat{p}')$.
Now, we take only the part of $\mathcal{F}(\hat{p})$ that corresponds to the positive frequencies and element-wise multiply it with the damping modes $\zeta$,
\begin{equation}
    \xi_j = \mathcal{F}^+_j(\hat{p}')\zeta_j, \quad j = 1,2,...,N_g
\end{equation}
In the next step, we take the inverse Fourier transform of the convolution. As a matter of fact, the training is done in batches, so the $x$ locations we are interested in calculating the effect of damping at may not necessarily collide with the spatial grid that we have previously constructed. So, we will do the inverse Fourier transform for these locations in the training batch. Ultimately, the damping term is found as
\begin{equation}
    \zeta p'(x) = \frac{1}{2K}\sum_{j = 1}^{N_g} \xi_j e^{ij\pi x} + \xi_j^* e^{-ij\pi x},
\end{equation}
where $\xi^*$ denotes the complex conjugate of $\xi$. We use the contribution of damping in the residual of the energy equation \eqref{eq:rijke_energy} when computing the physics-informed loss \eqref{eq:loss_energy}.

\subsection{Heat release}
Although the heat-release term acts as a Dirac delta in the spatial domain, the solver implements Galerkin decomposition, which projects this term onto a truncated set of modes. Plugging back the decomposition \eqref{eq:rijke_galerkin} in the left-hand side of the energy equation \eqref{eq:rijke_energy}, the remaining heat release term is equal to $\sum_{j = 1}^{N_g}2\dot{q}'\sin(j\pi x_f)\sin(j\pi x)$, which is the effective term in the simulations. As the number of Galerkin modes approaches to infinity, this approaches the Dirac delta. However, from a practical point of view, the number of modes in the simulation is finite, thus we use this expression to find the contribution of heat release in the residual of the energy equation \eqref{eq:rijke_energy} when computing the physics-informed loss \eqref{eq:loss_energy}.

\section{Effect of hyperparameter in the sine activation}\label{sec:hyp_a_appendix}
\begin{figure}[h!]
    \centering
    \includegraphics[width = \linewidth]{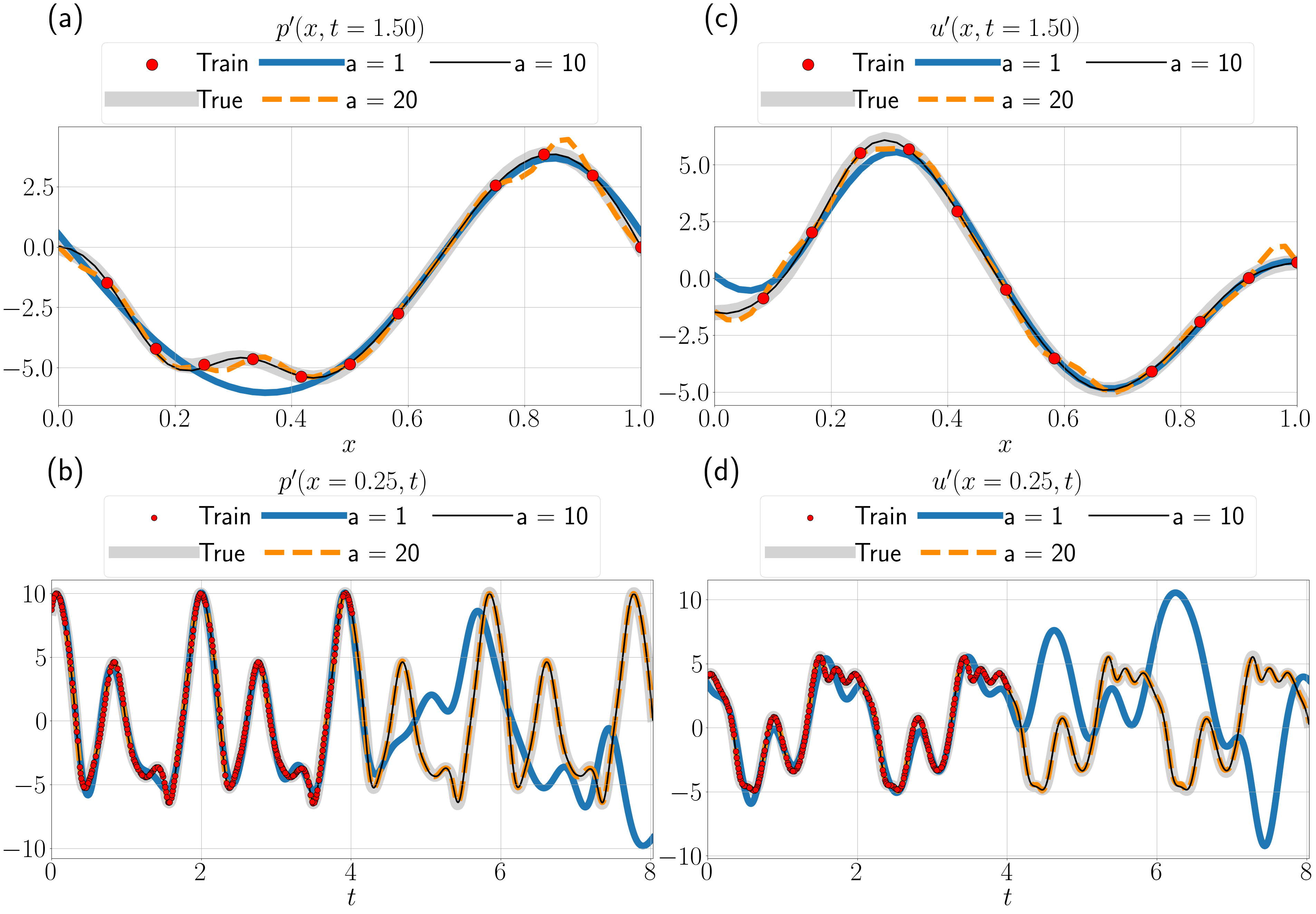}
    \caption{Effect of hyperparameter $a$ in the sine activation $\frac{1}{a}\sin(az)$}
    \label{fig:rijke_a}
\end{figure}
We observe that varying hyperparameter $a$ in the sine activation formulated as $\frac{1}{a}\sin(az)$ affects the frequency of the learned functions. Hence, it is a hyperparameter that requires tuning. Figure \ref{fig:rijke_a} illustrates this effect for the Rijke tube data discussed in Section \ref{sec:results_rijke_activation} for $a = 1, 10, 20$. Low $a$, $a = 1$, results in a low frequency model, whereas high, $a = 20$, results in a high frequency model, which can especially be observed in the pressure and velocity shapes in the spatial domain. We found the optimum value for this dataset as $a = 10$. Since the sampling frequency of the training data is high enough, we do not observe the effect of high $a$ in the time domain. \citet{Ziyin2020} reported similar findings in their studies when using the $1+\sin^2(z)$ activation.

\section{Travelling wave solution to the higher fidelity model}\label{sec:higher_fidelity_appendix}
In this model, the pressure and velocity are expressed as functions of two acoustic travelling waves, which propagate up- and downstream of the tube. These waves are derived by applying the method of characteristics to the acoustic wave equation and defined as $f$ and $g$, with propagation velocities $\tilde{\bar{c}}_1 \pm \tilde{\bar{u}}_1$ in the upstream region $\tilde{x} \leq \tilde{x}_f$; and $h$ and $j$, with propagation velocities $\tilde{\bar{c}}_2 \pm \tilde{\bar{u}}_2$ in the downstream region $\tilde{x} \geq \tilde{x}_f$. 
This model can also account for nonideal open boundary conditions via reflection coefficients $R_{in}$ and $R_{out}$ at the inlet and outlet, $f(\tilde{t}) = R_{in}g(\tilde{t}-\tilde{\tau}_u)$ and $j(\tilde{t}) = R_{out}h(\tilde{t}-\tilde{\tau}_d)$, where $\tilde{\tau_u}$ and $\tilde{\tau_d}$ are the travelling times of the waves from the flame to the up- and downstream boundaries, respectively. If the boundary conditions are ideal, i.e., fully reflective, then $R_{in}=R_{out}=-1$. The full set of equations that describe the dynamics of the waves are given by
\begin{equation}
\bm{X}
    \begin{bmatrix}
    g(\tilde{t}) \\ h(\tilde{t})
    \end{bmatrix}
    = 
 \bm{Y}  
    \begin{bmatrix}
    g(\tilde{t}-\tilde{\tau}_u) \\ h(\tilde{t}-\tilde{\tau}_d)
    \end{bmatrix}
    +
    \begin{bmatrix}
    0 \\ \frac{\tilde{\dot{q}}-\tilde{\bar{\dot{q}}}}{\tilde{A}_1\tilde{\bar{c}}_1}
    \end{bmatrix},
\end{equation}
where the matrices $\bm{X}$ and $\bm{Y}$ are functions of the mean-flow variables and are obtained from the jump conditions. We generate the data with the code implementation of the kinematic flame model from \cite{Aguilar2019}. (Code available at \cite{AguilarCode}).

\section{\rev{Training costs}}\label{sec:costs}
\rev{The training costs associated with some of the characteristic cases are provided below.}
\begin{table}[h!]
    \centering
    \caption{Computation time of twin experiments with 13 pressure sensors. Training with a batch size of 32 for 2000 epochs (on Intel Xeon(R) Gold 5218R CPU @ 2.10 Ghz x 80).}
    \begin{tabular}{c|c|c}
         Network name & Description & Training time \\ \hline 
         PI-P-GalNN & 10 Galerkin modes, 1 harmonic layer with 40 neurons & 1h 50m \\
         PI-GalNN & 10 Galerkin modes, 2 layers with 16 neurons  & 1h 50m \\
         PI-P-FNN & 3 layers with 32 neurons & 2h 40m
    \end{tabular}
    \label{tab:twin_cost}
\end{table}
\begin{table}[h!]
    \centering
    \caption{Computation time of nonideal boundary condition higher-fidelity experiments with 5 pressure sensors. Training with a batch size of 32 for 10000 epochs (on Intel Core(TM) i7-9750H CPU @  2.60 Ghz x 6).}
    \begin{tabular}{c|c|c}
         Network name & Description & Training time \\ \hline 
         PI-P-GalNN & 15 Galerkin modes, 1 harmonic layer with 10 neurons & 20m \\
         PI-P-FNN & 4 layers with 32 neurons & 35 
    \end{tabular}
    \label{tab:high_cost}
\end{table}

\clearpage

\section{\rev{Reconstruction from partial measurements with varying number of sensors}}\label{sec:partial_sensor}
\begin{figure*}[h!]
    \centering
    \includegraphics[width = \linewidth]{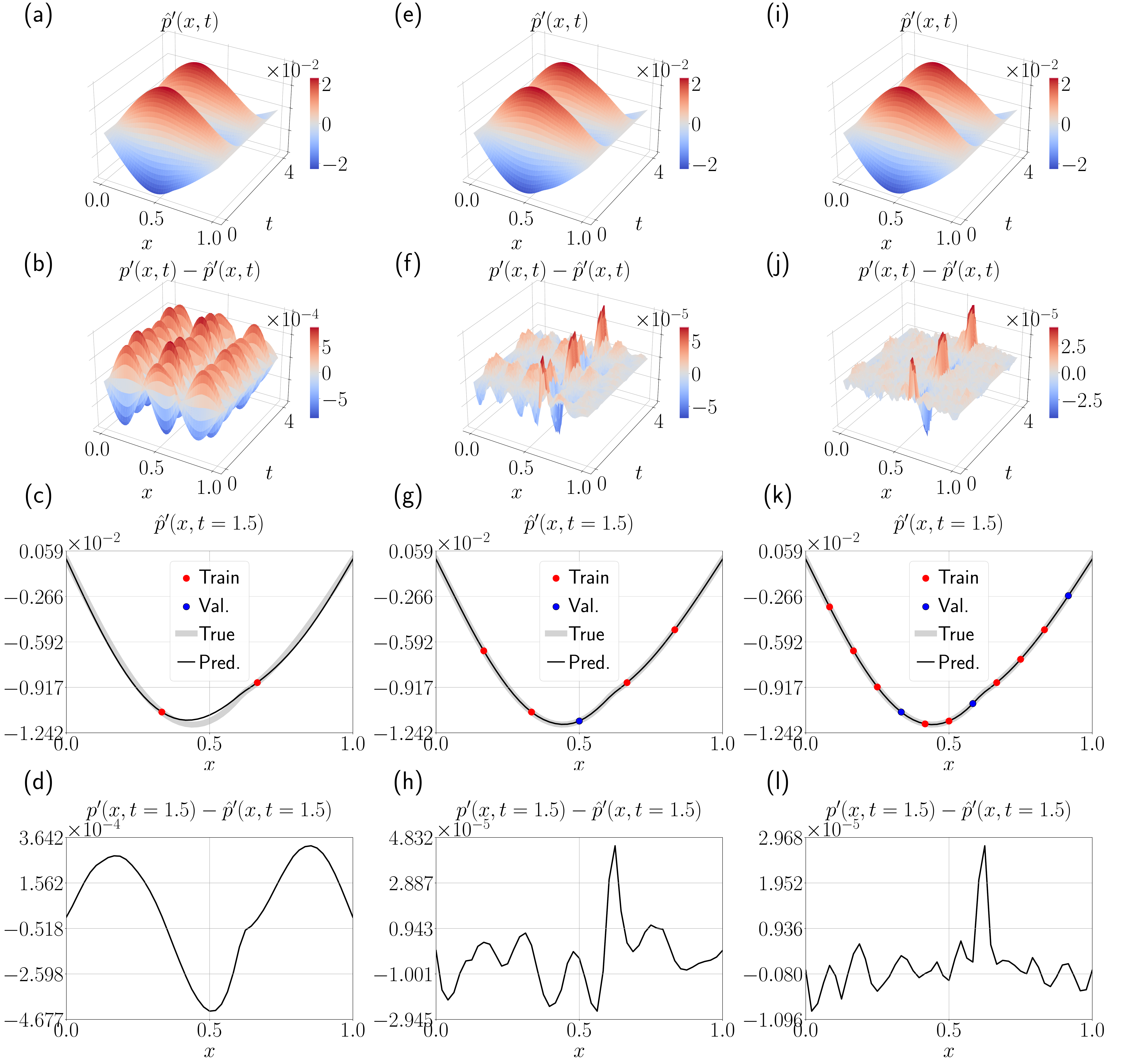}
    \caption{\rev{Reconstruction from only pressure measurements with physics-informed periodic GalNN. From left to right, (a-d) 2 sensors, 5 Galerkin modes, (e-h) 5 sensors, 10 Galerkin modes, (i-l) 11 sensors, 20 Galerkin modes. From top to bottom: acoustic pressure, error between ground truth and prediction, pressure prediction at a fixed time instance, and error at a fixed time instance, $t = 1.5$.}}
    \label{fig:sensor_grid_plot_P}
\end{figure*}

\begin{figure*}[t!]
    \centering
    \includegraphics[width = \linewidth]{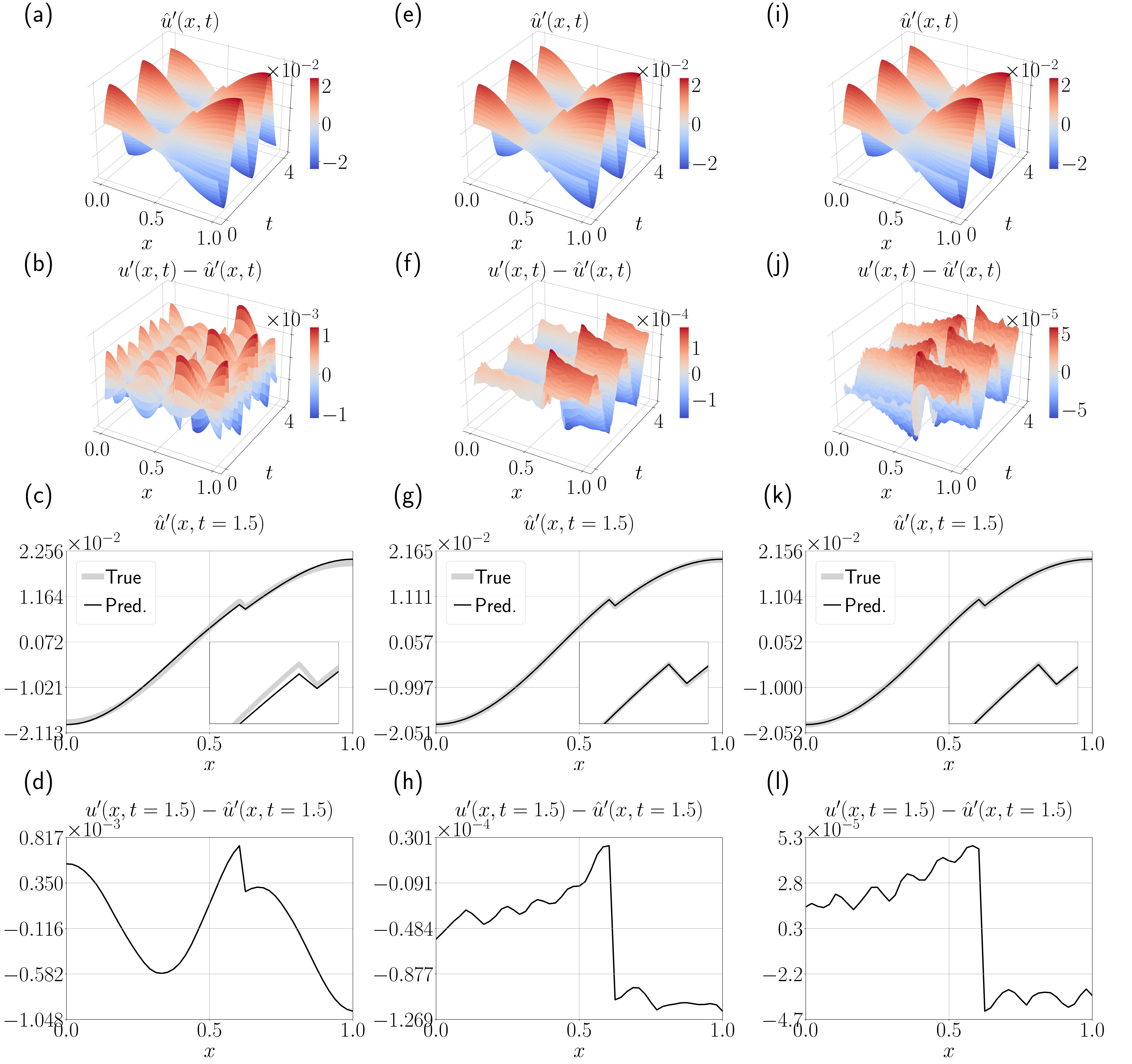}
    \caption{\rev{Reconstruction from only pressure measurements with physics-informed periodic GalNN. From left to right, (a-d) 2 sensors, 5 Galerkin modes, (e-h) 5 sensors, 10 Galerkin modes, (i-l) 11 sensors, 20 Galerkin modes. From top to bottom: acoustic velocity, error between ground truth and prediction, velocity prediction at a fixed time instance, and error at a fixed time instance, $t = 1.5$.}}
    \label{fig:sensor_grid_plot_U}
\end{figure*}
\clearpage
\bibliography{bibliography}
\end{document}